\documentclass[12pt]{article}
\usepackage{amsmath,amssymb}
\usepackage{amsmath}
\begin{document}

\begin{center}

{\LARGE {Accelerating internal dimensions and nonzero positive cosmological constant}}

\end{center}

\vskip 0.7cm

\renewcommand{\thefootnote}{\fnsymbol{footnote}}

\begin{center}
{\large Eun Kyung Park\footnote{E-mail: ekpark@pusan.ac.kr}$^{\rm a,b}$ and Pyung Seong Kwon\footnote{E-mail: bskwon@ks.ac.kr}$^{\rm a}$\\

\vskip 0.25cm}
\end{center}

\vskip 0.10cm

\begin{center}

{\it $^{\rm a}$Department of Energy Science, Kyungsung University, \\  Busan 48434, Korea\\

\vskip 0.3cm

$^{\rm b}$Department of Physics, Pusan National University,\\ Busan, 46241, Korea\\}

\end{center}

\thispagestyle{empty}

\vskip 2.0cm
\begin{center}
{\bf Abstract}
\vskip 0.1cm
\end{center}

We present a new scenario for the moduli stabilization with a very small but nonzero positive cosmological constant $\lambda$. In this scenario the complex structure moduli are still stabilized by the three-form fluxes as in the usual flux compactifications, but the K$\ddot{\rm a}$hler modulus is not fixed by the KKLT scenario. In our case the scale factor (or the K$\ddot{\rm a}$hler modulus) of the internal dimensions is basically allowed to change with time. But at the supergravity level it is fixed by a set of dynamical (plus constraint) equations defined on the 4D spacetime, not by the nonperturbative corrections of KKLT. Also at the supergravity level it is shown that $\lambda$ is fine-tuned to zero, $\lambda =0$, by the same set of 4D equations. This result changes once we admit $\alpha^{\prime}$-corrections of the string theory. The fine-tuning $\lambda=0$ changes into $\lambda = \frac{2}{3} Q$, where $Q$ is a constant representing quantum corrections of the brane and 6D action defined on the internal dimensions and its value is determined by the $\alpha^{\prime}$-corrections. It is also shown that this nonzero $\lambda$ must be positive and at the same time the internal dimensions must evolve with time almost at the same rate as the external dimensions in the case of nonzero $\lambda$.

\vskip 3cm

\vskip 0.25cm
\begin{center}
\end{center}

\newpage
\setcounter{page}{1}
\renewcommand{\thefootnote}{\number\value{footnote}}
\setcounter{footnote}{0}

\baselineskip 6.0mm

\vskip 1cm
\hspace{-0.65cm}{\bf \Large I. Introduction}
\vskip 0.5cm
\setcounter{equation}{0}
\renewcommand{\theequation}{1.\arabic{equation}}

It is known that the three-dimensional space of our present universe is now under accelerated expansion \cite{1}, which means that the
background vacuum of our present universe has its own energy density called dark energy, or the cosmological constant in the conventional sense.
The cosmological constant $\lambda$ is associated with quantum fluctuations of our vacuum and it must have some positive value to generate the
accelerated expansion described above. Indeed, observations show that $\lambda$ takes a positive value as mentioned above, but the mystery is
that it is unreasonably too small as compared with the theoretical value calculated from the quantum theory, and this leads to a hierarchy
problem called cosmological constant problem.

There have been many attempts to address this problem (for the review, see for instance \cite{2}), but it has remained an
unsolved problem. But very recently, a new mechanism has been proposed to address this problem \cite{3}, which is very distinguished from the
conventional theories where $\lambda$ is directly determined from the scalar potential ${\mathcal V}_{\rm scalar}$. In this mechanism $\lambda$
contains a supersymmetry breaking term ${\mathcal E}_{\rm SB}$ besides the usual ${\mathcal V}_{\rm scalar}$ of the $N=1$ supergravity and
where ${\mathcal E}_{\rm SB}$ has its own gauge arbitrariness. Thus the nonzero contributions to ${\mathcal V}_{\rm scalar}$ coming from the
perturbative and nonperturbative corrections, and also the NS-NS and R-R vacuum energies on the branes arising from quantum fluctuations are all gauged away by ${\mathcal E}_{\rm SB}$ (and by a certain self-tuning mechanism) and as a result $\lambda$ is fine-tuned to vanish. In this
self-tuning mechanism, whether $\lambda$ vanishes or not is basically determined by the tensor structure of ${\mathcal V}_{\rm scalar}$, not by the zero or nonzero values of ${\mathcal V}_{\rm scalar}$ itself. In \cite{3}, the above self-tuning mechanism has been applied to the well-known KKLT model \cite{4} to address the cosmological constant problem, especially aiming at explaining the vanishing $\lambda$ of our present universe.

In KKLT, the geometry (or the complex structure moduli) of the internal dimensions is stabilized by the three-form fluxes as in the usual flux
compactifications, but the scale factor (or the K$\ddot{\rm a}$hler modulus) of the internal dimensions is fixed by a certain KKLT mechanism in which the scalar potential acquires a minimum point by a K$\ddot{\rm a}$hler modulus-dependent nonperturbative correction. In the present paper we want to consider the self-tuning mechanism proposed in \cite{3} again. But this time we do not apply it to the KKLT. In the present paper we
assume that the complex structure moduli are still stabilized by the three-form fluxes. But the scale factor of the internal dimensions is not fixed by the KKLT scenario. In our present paper we basically assume that the internal dimensions are allowed to evolve with time. But nevertheless, we show that the scale factor of the internal dimensions is fixed at the supergravity level by a set of 4D equations, not by the K$\ddot{\rm a}$hler modulus-dependent nonperturbative corrections of KKLT, in the simplest setup. So in our model the no-scale structure is unbroken as in Ref. \cite{5}.

In this rather unconventional model $\lambda$ is fine-tuned to zero as in \cite{3}, again at the supergravity level. But once we admit $\alpha^{\prime}$-corrections of the string theory, the fine-tuning $\lambda =0$ changes into $\lambda = \frac{2}{3} Q$, where $Q$ is a constant representing quantum corrections of the brane and 6D action defined on the internal dimensions and its value is determined by the $\alpha^{\prime}$-corrections via another constant $c$. Namely $\lambda$ acquires nonzero values from the $\alpha^{\prime}$-corrections. In Sec. 7.2 we will show that this nonzero $\lambda$ must be positive and at the same time the internal dimensions must evolve with time almost at the same rate as the external dimensions in the case of nonzero $\lambda$. In this paper we aim at explaining both of these two aspects of $\lambda$ and the internal dimensions, based on the self-tuning mechanism presented in \cite{3}.

\vskip 1cm
\hspace{-0.65cm}{\bf \Large II. Time-dependent internal metric and $n$-form fields}
\vskip 0.5cm
\setcounter{equation}{0}
\renewcommand{\theequation}{2.\arabic{equation}}

\vskip 0.3cm
\hspace{-0.6cm}{\bf \large 2.1 Type IIB action}
\vskip 0.15cm

In the string frame the type IIB action is given by
\begin{equation}
 S_{\rm IIB}= \frac{1}{2 \kappa_{10}^2} \int d^{10}x \sqrt{-G_{10}} \Big( e^{-2 \Phi}
 \big( {\mathcal R}_{10} + 4 \, \partial_M \Phi \partial^M \Phi \big) - \frac{1}{2
 \cdot 3!} G_{(3)} \cdot {\bar G}_{(3)} - \frac{1}{4 \cdot 5!} {\tilde F}_{(5)}^2  \Big) \nonumber
\end{equation}
\begin{equation}
~~~~~~~~~~+ \frac{1}{8 i \kappa_{10}^2} \int e^{\Phi} A_{(4)} \wedge G_{(3)} \wedge {\bar G}_{(3)}  \,\,,
\end{equation}
where $G_{(3)} = F_{(3)} - i e^{-\Phi} H_{(3)}$ and ${\tilde F}_{(5)}$ is given by ${\tilde F}_{(5)} = F_{(5)} - \frac{1}{2} A_{(2)} \wedge
H_{(3)} + \frac{1}{2} B_{(2)} \wedge F_{(3)}$ with $F_{(n+1)} =dA_{(n)}$ etc. In (2.1), we have omitted the one-form field strength term
$F_{(1)}^2$ of the axion $A_{(0)}$ because, unlike the theories with scalar fields like quintessence, the axion does not play any
important role in our discussions of this paper. But in our paper we basically consider the case where the three-form fluxes  take nonzero values and the complex structure moduli of the internal dimensions are stabilized by these three-form fluxes. But this does not mean that we restrict our discussions only to the flux compactifications. Our discussions of this paper can be applied to both of the flux compactifications with $G_{(3)} \neq 0$ and the conventional compactifications with $G_{(3)}=0$.\footnote{There is a different viewpoint on the moduli stabilization which does not use the usual flux compactifications. For instance, in Sec. III of Ref. \cite{6} it was argued that the Calabi-Yau threefolds may be thought of as NS-NS solitons whose ADM masses are proportional to $1/{g_s^2}$. Hence in the limit $g_s \rightarrow 0$, these Calabi-Yau threefolds are very heavy and rigid and consequently deformations of internal geometry are highly suppressed.}

Now we introduce an ansatz for the 10D metric as
\begin{equation}
ds_{10}^2 =\alpha^2 ({\hat t}) e^{A(y)} {\hat g}_{\mu\nu} ({\hat x}) d{\hat x}^{\mu} d{\hat x}^{\nu} + \beta^2 ({\hat t}) e^{B(y)} {h}_{mn} (y)
d {y}^m d {y}^n \,\,,
\end{equation}
where ${\hat g}_{\mu\nu} ({\hat x})$ is the metric of the 4D spacetime,
\begin{equation}
{\hat g}_{\mu\nu} ({\hat x}) d{\hat x}^{\mu} d{\hat x}^{\nu}= - d {\hat t}^2 + a^2 ({\hat t}) d {\vec x}_3^2 \,\,,
\end{equation}
while ${h}_{mn} (y)$ represents the metric of the 6D internal dimensions. In (2.2), $\alpha^2 ({\hat t})$ is an extra degree of freedom which
could have been absorbed into ${\hat g}_{\mu\nu} ({\hat x}) d{\hat x}^{\mu} d{\hat x}^{\nu}$ by the coordinate transformation $d {\hat t}
\rightarrow dt \equiv \alpha({\hat t}) d {\hat t}$, so it can be taken arbitrarily as we wish. Similarly, $e^{B(y)}$ is also an extra degree of
freedom which can be taken arbitrarily as we wish. So we will take $\alpha ({\hat t})$ and $B(y)$ properly in the metric (2.2) later.

The metric (2.2) contains the time-dependent scale factor $\beta^2 ({\hat t})$ for the internal dimensions, which means that the internal
dimensions are basically allowed to evolve with time and this is one of the main points of our discussion distinguished from the usual
higher-dimensional theories in which the volume of the internal space is fixed by $\beta^2 ({\hat t})=1$ from the beginning. Since the metric of the
internal space changes with time, we may have to allow the time-dependence of the other fields as well. We introduce an ansatz for the dilaton as
\begin{equation}
e^{\Phi (y,{\hat t})} = g_s  \gamma({\hat t}) e^{\Phi_S (y)} \,\,,
\end{equation}
where $g_s$ is the string coupling constant and vacuum expectation value of $e^{\Phi_S (y)}$ is equal to one, $< e^{\Phi_S}> =1$. Similarly, the ansatz for the R-R five-form $\tilde{F}_{(5)}$ and the three-form $G_{(3)}$ are given respectively by
\begin{equation}
\tilde{F}_{(5)} = \sigma ({\hat t}) (1+ \ast_{10}) d\xi (y) \wedge \sqrt{-{\hat g}_4} \,d{\hat t} \wedge dx^1 \wedge dx^2 \wedge dx^3 \,\,, \nonumber
\end{equation}
\begin{equation}
\longrightarrow ~\,\, A_{(4)} = \sigma ({\hat t}) \xi (y) \sqrt{-{\hat g}_4} \, d {\hat t} \wedge dx^1 \wedge dx^2 \wedge dx^3 \,\,, ~~~~~~~~
\end{equation}
where ${\hat g}_4$ is the determinant of ${\hat g}_{\mu\nu}$ and therefore $\sqrt{-{\hat g}_4} = a^3 ({\hat t})$, and
\begin{equation}
F_{(3)} = \eta({\hat t})\, {\mathcal F}_{(3)}(y)\,\,,~~~ H_{(3)} = \eta ({\hat t}) \,{\mathcal H}_{(3)}(y)~~~~\rightarrow~~~~ G_{(3)} = \eta(\hat{t})\, {\mathcal G}_{(3)}(y)\,\,,
\end{equation}
where ${\mathcal G}_{(3)}(y) \equiv {\mathcal F}_{(3)} - i \widetilde{{{\rm Im} \tau}}\, {\mathcal H}_{(3)}$ with $ \widetilde{{\rm Im} \tau} \equiv (g_s \gamma \,({\hat t})\, e^{\Phi_S})^{-1}$.

Upon reduction (2.2), and taking $B(y) = \Phi_S (y) - A(y)$, one finds that (2.1) now reduces to
\begin{equation}
 S_{\rm IIB} = \frac{1}{2 \kappa_{10}^2 g_s^2} \Big( \int d^4 {\hat x} \sqrt{-{\hat g}_4} \frac{\alpha^4 \beta^4}{\gamma^2} \Big) \Big(\int d^6
 y \sqrt{h_6} ({\mathcal R}_6 ({h}_{mn}) - 2 {\mathcal H}) \Big) ~~~~~~~~~~~~~~~~~~~~\nonumber
\end{equation}
\begin{equation}
+\frac{1}{2 \kappa_{10}^2 g_s^2} \Big( \int d^4 {\hat x} \sqrt{-{\hat g}_4} \frac{\sigma^2 \beta^4}{\alpha^4} \Big)\Big(\int d^6 y \sqrt{h_6}
\frac{g_s^2}{2} e^{2\Phi_S -4A} (\partial \xi)^2 \Big) ~~~\,\nonumber
\end{equation}
\begin{equation}
- \frac{1}{2 \kappa_{10}^2 g_s^2} \Big( \int d^4 {\hat x} \sqrt{-{\hat g}_4} \alpha^4 \eta^2 \Big) \Big(\int d^6 y \sqrt{h_6}\, \frac{g_s^2}{3!}
\, e^{2A} {\mathcal G}_{mnp}^{+} {\bar {\mathcal G}}^{+mnp} \Big) ~~\,\nonumber
\end{equation}
\begin{equation}
+ \frac{1}{2 \kappa_{10}^2 g_s^2} \Big( \int d^4 {\hat x} \sqrt{-{\hat g}_4} \frac{\alpha^2 \beta^6}{\gamma^2} \hat{{\mathcal R}}_4^{(\rm eff)} ({\hat g}_{\mu\nu}, \alpha, \beta, \gamma) \Big) \Big(\int d^6 y \sqrt{h_6} \,  e^{\Phi_S -2A} \Big) \nonumber
\end{equation}
\begin{equation}
 +\,{\rm topological ~terms}  \,\,,
\end{equation}
where $\mathcal{H} \equiv \frac{1}{2} (\partial \Phi_S )^2 - (\partial \Phi_S ) (\partial A ) + (\partial A )^2$ and ${\mathcal G}_{mnp}^{+}$
represents the imaginary anti self dual (IASD) peace of the ${\mathcal G}_{mnp}$, ${\mathcal G}_{(3)}^{+} \equiv {\mathcal G}_{(3)}^{\rm IASD}$. Also, $\hat{{\mathcal R}}_4^{(\rm
eff)} ({\hat g}_{\mu\nu}, \alpha, \beta, \gamma)$ and topological terms are given, respectively, as follows.

First, $\hat{{\mathcal R}}_4^{(\rm eff)} ({\hat g}_{\mu\nu}, \alpha, \beta, \gamma)$ represents
\begin{equation}
\hat{{\mathcal R}}_4^{(\rm eff)} ({\hat g}_{\mu\nu}, \alpha, \beta, \gamma) = \hat{{\mathcal R}}_4 ({\hat g}_{\mu\nu}) + 6 \frac{\ddot{\alpha}}{\alpha} +
12\frac{\ddot{\beta}}{\beta} + 30 \Big( \frac{\dot{\beta}}{\beta} \Big)^2 + 18 \frac{\dot{a}}{a} \frac{\dot{\alpha}}{\alpha} +
36\frac{\dot{a}}{a} \frac{\dot{\beta}}{\beta} + 24 \frac{\dot{\alpha}}{\alpha} \frac{\dot{\beta}}{\beta}  - 4 \Big( \frac{\dot{\gamma}}{\gamma}
\Big)^2 \,\,,
\end{equation}
where the "dot" denotes the derivative with respect to ${\hat t}$ and $\hat{{\mathcal R}}_4 ({\hat g}_{\mu\nu})$ is the usual Ricci-scalar of the 4D
metric (2.3):
\begin{equation}
\hat{{\mathcal R}}_4 ({\hat g}_{\mu\nu}) = 6 \Big( \frac{\ddot{a}}{a}  +\Big(\frac{\dot{a}}{a}\Big)^2 \Big)\,\,.
\end{equation}
The above $\hat{{\mathcal R}}_4^{(\rm eff)} ({\hat g}_{\mu\nu}, \alpha, \beta, \gamma)$ reduces to $\hat{{\mathcal R}}_4 ({\hat g}_{\mu\nu})$ in the
time-independent limit $\alpha ({\hat t})= \beta ({\hat t})= \gamma ({\hat t})=1$. Also one can show that (2.8) can be rewritten as
\begin{equation}
\sqrt{-{\hat g}_4} \frac{\alpha^2 \beta^6}{\gamma^2} \hat{{\mathcal R}}_4^{(\rm eff)}({\hat g}_{\mu\nu}, \alpha, \beta, \gamma) = \frac{d}{d{\hat t}} \Bigg[ a^3 \Big( \frac{\alpha^2 \beta^6}{\gamma^2} \Big) \Bigg( 6 \Big(\frac{\dot{a}}{a} + \frac{\dot{\alpha}}{\alpha} \Big) +12 \frac{\dot{\beta}}{\beta}
\Bigg) \Bigg] ~~~~~~~~~~~~~~~~~~~~~~~~~~~~~~~~~~~~~~~~\nonumber
\end{equation}
\begin{equation}
~~~~~+ a^3 \Big( \frac{\alpha^2 \beta^6}{\gamma^2} \Big) \Bigg(- 6 \Big(\frac{\dot{a}}{a} + \frac{\dot{\alpha}}{\alpha} \Big)^2 +12\Big(\frac{\dot{a}}{a} + \frac{\dot{\alpha}}{\alpha} \Big) \Big( \frac{\dot{\gamma}}{\gamma} - 3 \frac{\dot{\beta}}{\beta} \Big) - 4 \Big( \frac{\dot{\gamma}}{\gamma} -
3 \frac{\dot{\beta}}{\beta} \Big)^2 + 6 \Big(\frac{\dot{\beta}}{\beta}\Big)^2 \Bigg) \,\,.
\end{equation}

The topological terms, on the other hand, are given by
\begin{equation}
{\rm topological~\,terms} = \frac{i}{4 \kappa_{10}^2} \Big( \int d^4 {\hat x} \sqrt{-{\hat g}_4} \,\alpha^4 \eta^2 \Big)\Big( \frac{1}{ {\rm Im}
\tau} \int \frac{e^{2A-\Phi_S}}{g_s} \, {\mathcal G}_{(3)} \wedge {\bar {\mathcal G}}_{(3)} \Big) \nonumber
\end{equation}
\begin{equation}
~~~~~~~~~~~~\,~~~~- \frac{i}{4 \kappa_{10}^2} \Big( \int d^4 {\hat x} \sqrt{-{\hat g}_4} \,\sigma \gamma \eta^2 \Big)\Big( \frac{1}{ {\rm Im}
\tau} \int  \xi \, {\mathcal G}_{(3)} \wedge {\bar {\mathcal G}}_{(3)} \Big) \,\,,
\end{equation}
where $({ {\rm Im} \tau})^{-1} \equiv g_s \, e^{\Phi_S}$. In (2.11), the second term is just the Chern-Simons term $\int e^{\Phi} A_{(4)} \wedge
G_{(3)} \wedge {\bar G}_{(3)}$. But the first term comes from the $G_{(3)} \cdot {\bar G}_{(3)}$ term of the action (2.1). Using the identity
\begin{equation}
{\mathcal G}_{(3)} \wedge {\ast_6} {\bar {\mathcal G}}_{(3)}= -i {\mathcal G}_{(3)} \wedge {\bar {\mathcal G}}_{(3)} + 2i {\mathcal G}_{(3)}^+ \wedge {\bar {\mathcal G}}_{(3)}^+ \,\,,
\end{equation}
together with ${\ast_6} {\mathcal G}_{(3)}^+ = -i {\mathcal G}_{(3)}^+$, one can show that the $G_{(3)} \cdot {\bar G}_{(3)}$ term in (2.1) can
be decomposed as
\begin{equation}
- \frac{1}{24 \kappa_{10}^2} \int d^{10}x \sqrt{-G_{10}} \, G_{(3)} \cdot {\bar G}_{(3)} = \Big( \int d^4 {\hat x} \sqrt{-{\hat g}_4} \,\alpha^4 \eta^2 \Big)\Big( \frac{i}{4 \kappa_{10}^2} \frac{1}{ {\rm Im} \tau} \int  \frac{e^{2A-\Phi_S}}{g_s} \, {\mathcal G}_{(3)} \wedge {\bar
{\mathcal G}}_{(3)} \Big) \nonumber
\end{equation}
\begin{equation}
~~~~~~~~~~~~~~~~~~- \Big( \int d^4 {\hat x} \sqrt{-{\hat g}_4} \,\alpha^4 \eta^2 \Big)\Big( \frac{1}{12 \kappa_{10}^2} \int d^6 y \sqrt{h_6}\,
e^{2A} {\mathcal G}_{mnp}^{+} {\bar {\mathcal G}}^{+mnp} \Big) \,\,,
\end{equation}
and these two terms become, respectively, the first term of (2.11) and the third term of (2.7).

Before we finish this section we say a few words about the ansatz (2.6). In the usual time-independent theories the three-form fields $F_{(3)}$ and $H_{(3)}$ are quantized in such a way that each of their integrations along a certain basis of three-cycles in the internal dimensions becomes a constant proportional to an integer
\begin{equation}
\frac{1}{2 \pi \alpha^{\prime}} \int F_{(3)} \in 2 \pi {\mathbb{Z}} \,\,, ~~~~~\frac{1}{2 \pi \alpha^{\prime}} \int H_{(3)} \in 2 \pi {\mathbb{Z}} \,\,.
\end{equation}
But in our case (2.14) cannot be satisfied because $F_{(3)}$ and $H_{(3)}$ on the left hand sides have now time dependences by (2.6). Besides this, we have the Bianchi identity for $\tilde{F}_{(5)}$
\begin{equation}
d \tilde{F}_{(5)} = H_{(3)} \wedge F_{(3)} + 2 \kappa^2_{10} \mu_0 \rho_3^{\rm loc} \,\,.
\end{equation}
This equation, under integration over the internal manifold $\mathcal{M}$, leads to the tadpole-cancellation condition
\begin{equation}
\frac{1}{2 \kappa^2_{10} T_0} \int_{\mathcal{M}} H_{(3)} \wedge F_{(3)} + Q_3^{\rm loc} =0 \,\,,~~~(T_0 = \mu_0 )\,\,,
\end{equation}
where $Q_3^{\rm loc}$ is the total D3 charge associated with $\rho_3^{\rm loc}$. In (2.16) the total charge $Q_3^{\rm loc}$ is a constant, thus the first term must also be a constant. However, this is also impossible because $F_{(3)}$ and $H_{(3)}$ are both time-dependent by (2.6) again in our case. The flux quantizations in (2.14) play a crucial role in the stabilization of the complex structure (and the dilaton) moduli and the tadpole condition (2.16) also needs to be satisfied for a consistency of the quantum theory. In Sec. 3.3 we will be back to this point to show that the inconsistencies described above are virtually nonexistent.

\vskip 0.3cm
\hspace{-0.6cm}{\bf \large 2.2 Brane action}
\vskip 0.15cm

In addition to the action $S_{\rm IIB}$, we also have local terms
\begin{equation}
S_{\rm brane} = - \int d^4 {\hat x} \sqrt{-det \,(G_{\mu\nu})}\,\, T(\Phi) + \mu (\Phi) \int A_{(4)} \,\,,
\end{equation}
where $G_{\mu\nu}$ is a pullback of the target space metric $G_{MN}$ to the 4D brane world. In (2.17), $T(\phi)$ represents the tension of the
D3-brane; it is given by $T(\Phi) = T_0 \,e^{-\Phi}$ at the tree level, but $T(\Phi) = T_0 \,e^{-\Phi} + \rho_{\rm vac} (\Phi)$ at the quantum level, where $\rho_{\rm vac} (\Phi)$ represents quantum correction terms; $\rho_{\rm vac} (\Phi) = \sum_{n=0}^{\infty} T_{n+1} e^{n \Phi}$
(see, for instance, Ref. \cite{7}). So $T(\Phi)$ becomes
\begin{equation}
T(\Phi) = T_0\, e^{-\Phi} \Gamma_{NS} (\Phi) \,\,,
\end{equation}
where
\begin{equation}
\Gamma_{NS} (\Phi) = 1 + \sum_{n=1}^{\infty} {\hat T}_{n}\, e^{n \Phi}\,\,,~~~~~({\hat T}_{n} \equiv \frac{T_n}{T_0})\,\,.
\end{equation}
Similarly, $\mu (\Phi)$ is given by $\mu (\Phi) = \mu_0$ at the tree level, but it turns into $\mu (\Phi)=\mu_0 + \delta \mu (\Phi)$ at the
quantum level where $\delta \mu (\Phi)$ is given by $\delta \mu (\Phi) =  \sum_{n=1}^{\infty} \mu_{n} e^{n \Phi}$. So $\mu (\Phi)$ becomes
\begin{equation}
\mu (\Phi)=\mu_0 \Gamma_R (\Phi) \,\,,
\end{equation}
where
\begin{equation}
\Gamma_{R} (\Phi) = 1 + \sum_{n=1}^{\infty} {\hat \mu}_{n}\, e^{n \Phi}\,\,,~~~~~\Big({\hat \mu}_n \equiv \frac{\mu_n}{\mu_0}\Big)\,\,.
\end{equation}

Using (2.5) together with (2.18) and (2.20), one finds that (2.17) reduces to
\begin{equation}
S_{\rm brane} = \int d^4 {\hat x} \sqrt{-{\hat g}_4} \int d^6 y \sqrt{h_6} \Big( - \frac{T_0}{g_s} \chi^{{1}/{2}}(y) \frac{\alpha^4}{\gamma} \Gamma_{NS} + \mu_0 \xi(y) \sigma \Gamma_R  \,\Big) \delta^6 (y) \,\,,
\end{equation}
where the 6D delta function is normalized by $\int d^6 y \sqrt{h_6} \delta^6 (y)=1$ and $\chi$ is defined by
\begin{equation}
\chi = e^{4A-2\Phi_S} \,\,.
\end{equation}
In (2.22), the first term constitutes the NS-NS part of the action, while the second term is an R-R counterpart of the first term. For the BPS-branes ($T_0 = \mu_{0}$) these two terms cancel out at the tree level, which is related with the fact that D3-brane potential defined by (see Sec. 7 of \cite{3})
\begin{equation}
\frac{1}{g_s} \Phi_{-} \equiv \frac{\chi^{1/2}}{g_s} - \xi
\end{equation}
vanishes in the imaginary self dual (ISD) backgrounds. Indeed, these two terms are expected to cancel out to all orders of perturbations when
supersymmetry of the brane region is unbroken. But in (2.22), such a cancellation cannot be achieved unless the time-dependent factor
$\frac{\alpha^4}{\gamma}$ of the first term coincides with $\sigma$ of the second term. So we choose $\alpha({\hat t})$ as
\begin{equation}
\alpha^4 = \sigma \gamma \,\,,
\end{equation}
so that the cancelation occurs for the BPS-branes.

\vskip 1cm
\hspace{-0.65cm}{\bf \Large III. 6D effective action and moduli stabilization}
\vskip 0.5cm
\setcounter{equation}{0}
\renewcommand{\theequation}{3.\arabic{equation}}

Now we turn to the equation of motion for $\xi(y)$. We have important reasons to find this equation of motion. In the self-tuning mechanism of \cite{3} (and therefore in the self-tuning mechanism of this paper) $\lambda$ contains the action densities ${\hat S}_{\rm brane}$ and ${\hat S}_{\rm topological}$ (see Eq. (5.7)) which are defined respectively by (4.5) and (3.10). In Sec. 6.2 we want to show that these action densities both vanish at the tree level. But at the tree level these action densities are proportional to the D3-brane potential $\Phi_{-} (y)$ defined in (2.24)(see (6.10) and (6.11)). So if we can show that $\Phi_{-} (y)$ vanishes at the tree level, then we can say that ${\hat S}_{\rm brane}$ and ${\hat S}_{\rm topological}$ in $\lambda$ both vanish at the tree level.

In Sec. 3.1 we will obtain time-independent equation of motion for $\xi (y)$ starting from the time-dependent 10D Lagrangian. This equation of motion is important itself because it will be used to show that $\Phi_{-} (y)$ really vanishes at the tree level (see Sec. 6.2). Also, equation of motion for $\xi (y)$ leads to the time-independent equation of motion for $\tilde{F}_{(5)}$ and Bianchi identity, which then solve the inconsistency problems stated at the end of Sec. 2.1.

More importantly, during the process of obtaining the time-independent equation for $\tilde{F}_{(5)}$ (and time-independent brane action $S_{\rm brane}$), we will find a couple of constraint equations which relates the time-dependent factors $\eta (t)$ and $\sigma (t)$ to the remaining moduli $\alpha (t)$, $\beta (t)$ and $\gamma (t)$. These relations then enable us to obtain a time-independent action defined on the internal sector (see Sec. 3.2) and furthermore, they relate the whole physical quantities and equations like superpotential, scalar potential and equations for the moduli stabilizations etc. of this paper to the corresponding physical quantities and equations of the existing time-independent theories, as we will see in the followings. Now we start our discussion from the equation of motion for $\xi (y)$.

\vskip 0.3cm
\hspace{-0.6cm}{\bf \large 3.1 Time-independent equation of motion for $\xi(y)$}
\vskip 0.15cm

From (2.7) and (2.22), the 10D Lagrangian for $\xi (y)$ can be written as
\begin{equation}
2 \kappa_{10}^2 L_{\xi (y)} = \frac{1}{2} \sqrt{-{\hat g}_4}\,\sqrt{h_6}\, \frac{\sigma^2 \beta^4}{\alpha^4} \chi^{-1} (\partial \xi)^2 +
\frac{i}{12} \sqrt{-{\hat g}_4}\,\sqrt{h_6}\, \sigma \gamma \eta^2 \frac{\xi}{ {\rm Im} \tau}\,{\mathcal G}_{mnp} \big(*_6 {\bar {\mathcal G}}
\big)^{mnp} \nonumber
\end{equation}
\begin{equation}
+ 2 \kappa_{10}^2 \sqrt{-{\hat g}_4}\,\sqrt{h_6}\, \mu_0 \sigma \Gamma_R \xi(y) \delta^6 (y) \,\,,~~~~~~~~~~~~~~~~~~~~~~~~~~~
\end{equation}
(where the second term comes from the topological term in (2.11)) and from this Lagrangian we obtain the equation of motion
\begin{equation}
\frac{1}{\sqrt{h_6}} \partial^m \Big( \sqrt{h_6} \chi^{-1} h_{mn} (\partial^n \xi) \Big) =
\frac{i}{12}\frac{\alpha^4}{\beta^4}\frac{\gamma}{\sigma} \eta^2 \,\frac{1}{{\rm Im} \tau} {\mathcal G}_{mnp} \big(*_6 {\bar {\mathcal G}}
\big)^{mnp} + 2 \kappa_{10}^2 \mu_0 \frac{\alpha^4}{\beta^4}\,\frac{\Gamma_R}{\sigma} \delta^6 (y) \,\,.
\end{equation}
(3.2) differs from the corresponding equation of the time-independent theory in \cite{3}. The left hand side is independent of ${\hat t}$ as in
\cite{3}. But each term on the right hand side contains extra factors $\frac{\alpha^4 \gamma}{\beta^4 \sigma}\, \eta^2$ and $\frac{\alpha^4
\Gamma_R}{\beta^4 \sigma}$ respectively, which are functions of ${\hat t}$. So in order that the equality holds we must require that the functions $\frac{\alpha^4 \gamma}{\beta^4 \sigma}\, \eta^2$ and $\frac{\alpha^4 \Gamma_R}{\beta^4 \sigma}$ must be constants.
We set
\begin{equation}
\frac{\alpha^4}{\beta^4} \frac{\gamma}{\sigma} \eta^2 =1 \,\,,
\end{equation}
and similarly
\begin{equation}
\frac{\alpha^4}{\beta^4}\,\frac{\Gamma_R}{\sigma} =1 \,\,,
\end{equation}
where $\Gamma_R$ represents the values of $\Gamma_R (\Phi (y, {\hat t}\,))$ at $y=0$ by the Dirac delta $\delta^6 (y)$, so it is only a function of ${\hat t}$. By (3.3) and (3.4), (3.2) reduces to the time-independent equation
\begin{equation}
\nabla^2 \xi = \frac{i}{12 {\rm Im} \tau} \,\chi \, {\mathcal G}_{mnp} \big(*_6 {\bar {\mathcal G}} \big)^{mnp} + 2 \chi^{-1/2} ({\partial
\chi^{1/2}})(\partial \xi) + 2 \kappa^2_{10} \mu_0 \chi \delta^6 (y) \,\,.
\end{equation}
Equation (3.5) will be used in Sec. 6.2 to show that ${\hat S}_{\rm brane}$ and ${\hat S}_{\rm topological}$ vanish at the tree level as mentioned in the introduction of this section.

\vskip 0.3cm
\hspace{-0.6cm}{\bf \large 3.2 6D effective action}
\vskip 0.15cm

The time-independent equation (3.5) can be obtained directly from a 6D effective action defined on the 6D internal sector. From (2.25) and (3.3) one finds that
\begin{equation}
\eta = \frac{\beta^2}{\gamma}\,\,,
\end{equation}
and using (2.25) and (3.6) one obtains a time-independent 6D action from (2.7) and (2.11) :
\begin{equation}
S_{\rm IIB} / \Big( \int  d^4 {\hat x} \sqrt{-{\hat g}_4}\, \frac{\alpha^4  \beta^4}{\gamma^2} \Big) = \frac{1}{2 \kappa_{10}^2 g_s^2} \int d^6
{y} \sqrt{h_6} \Big({\mathcal R}_6 (h_{mn}) - {\mathcal L}_F + c \chi^{-1/2} \Big) \nonumber
\end{equation}
\begin{equation}
~~~~~~~~~~+ {\hat S}_{\rm topological}\,\,,
\end{equation}
where ${\mathcal L}_F$ and $c$ are defined by
\begin{equation}
{\mathcal L}_F = 2 {\mathcal H} - \frac{g_s^2}{2} \chi^{-1} (\partial \xi)^2 + \frac{g_s^2}{3!} \, e^{2A} \, {\mathcal G}_{mnp}^+ {\bar
{\mathcal G}}^{+mnp}\,\,,
\end{equation}
and
\begin{equation}
 c= \frac{ \int d^4 {\hat x} \sqrt{-{\hat g}_4} \,\frac{\alpha^2 \beta^6}{\gamma^2} \,{\hat{\mathcal R}}_4^{(\rm eff)}({\hat g}_{\mu\nu}, \alpha, \beta, \gamma)}{\int d^4 {\hat x} \sqrt{-{\hat g}_4} \,\frac{\alpha^4 \beta^4}{\gamma^2} } \,\,,
\end{equation}
while the topological term ${\hat S}_{\rm topological}$ is given by
\begin{equation}
{\hat S}_{\rm topological} = \frac{i}{4 \kappa^2_{10} {\rm Im} \tau} \int (\frac{\chi^{1/2}}{g_s} - \xi) \, {\mathcal G}_{(3)} \wedge {\bar {\mathcal G}}_{(3)} \,\,.
\end{equation}
The action (3.7) coincides with the corresponding action in Ref. \cite{3} only except that $\beta$ is replaced by a new constant $c$ (see Eq. (3.7) of Ref. \cite{3}). (3.7) defines a time-independent theory at least on the 6D internal sector. The field contents in (3.7) are defined with support on the 6D internal space, and field equations obtained from (3.7) are all time-independent. Indeed one can show that the time-independent equation (3.5) can be obtained from (3.7) as mentioned at the beginning of Sec. 3.2.

The 6D Einstein equation defined on the internal dimensions also can be obtained from (3.7). Varying (3.7) with respect to $\delta h^{mn}$ one obtains
\begin{equation}
{\mathcal R}_{mn} -\frac{1}{2} h_{mn} {\mathcal R}_6 - \frac{1}{2} T_{mn} - \frac{c}{2} \chi^{-1/2} h_{mn} =0 \,\,,
\end{equation}
where the energy-momentum tensor $T_{mn}$ is defined by
\begin{equation}
T_{mn} = \frac{2}{\sqrt{h_6}} \frac{\delta S_F}{\delta h^{mn}}\,\,,~~~~~~~~~~(S_F \equiv \int d^6 y \sqrt{h_6} {\mathcal L}_F \,)\,\,.
\end{equation}
This Einstein equation will be used in Sec. V to obtain a self-tuning equation for $\lambda$, which is the essential part of this paper. In (3.11), we do not take ${\mathcal R}_{mn}={\mathcal R}_6 =0$. In our discussions of this paper we basically consider the Calabi-Yau (CY) compactifications. However, the self-tuning mechanism proposed in this paper (and in \cite{3}) is not the restricted mechanism designed only for the type IIB theory. It is a general formalism which can be applied to any types of flux compactifications. For instance, in the heterotic compactifications the internal manifold is no longer a K$\ddot{\rm a}$hler once the $H_{(3)}$-fluxes are turned on. In this case the background flux $H_{(3)}$ is given by $H_{(3)} = - \frac{i}{2} (\partial - {\bar{\partial}}) J$ \cite{7-1}, which means that $dJ=0$ cannot be satisfied when $H_{(3)}$ is nonzero. In general, the zeroth order CY metric becomes non-K$\ddot{\rm a}$hler at the leading order $\alpha^{\prime}$-corrections of the heterotic theory. Namely the metric acquires the correction terms (see Eq.(2.13) of Ref. \cite{8}, for instance),
\begin{equation}
h_{mn}=h_{mn}^{(0)} + h_{mn}^{(1)} + h_{mn}^{(2)} + \cdots = h_{mn}^{(0)} +\delta_Q h_{mn}
\end{equation}
at the quantum level, and in this case, ${\mathcal R}_{mn} (h_{mn})$ and ${\mathcal R}_{6} (h_{mn})$ do not vanish off-shell, though we have ${\mathcal
R}_{mn} (h_{mn}^{(0)})={\mathcal R}_{6} (h_{mn}^{(0)})=0$.\footnote{In fact, though we do not take ${\mathcal R}_{mn} (h_{mn}) = {\mathcal R}_{6} (h_{mn}) =0$, these nonzero Ricci tensor and Ricci scalar do not appear in the resulting equations. They cancel themselves out during the process of obtaining the resulting equations anyway.}

\vskip 0.3cm
\hspace{-0.6cm}{\bf \large 3.3 Equation of motion for $\tilde{F}_{(5)}$ and Bianchi identity}
\vskip 0.15cm

Equation (3.5) is the equation of motion for $\tilde{F}_{(5)}$. Using
\begin{equation}
\tilde{\mathcal F}_{(5)} = (1+ \ast_{10}) d\xi (y) \wedge \sqrt{-{\hat g}_4} \,d{\hat t} \wedge dx^1 \wedge dx^2 \wedge dx^3 \,\,,
\end{equation}
one can show that (3.5) is equivalent (see Sec. 7.2 of Ref.\cite{3}) to
\begin{equation}
d \ast_{10} \tilde{\mathcal F}_{(5)} = - \frac{{\mathcal G}_{(3)} \wedge {\bar{\mathcal G}}_{(3)}}{2i {\rm Im} \tau} + 2 \kappa^2_{10} \mu_0 \rho_3^{\rm loc} \,\,,
\end{equation}
where $\tilde{\mathcal F}_{(5)}$ is time-independent five-form defined by $\tilde{\mathcal F}_{(5)} \equiv \tilde{F}_{(5)} / \sigma (\hat t)$. Equation (3.15) is the usual time-independent equation for the five-form $\tilde{F}_{(5)}$. $\tilde{\mathcal F}_{(5)}$ and ${\mathcal G}_{(3)}$ in (3.15) play exactly the same role as the usual $\tilde{F}_{(5)}$ and $G_{(3)}$ of the time-independent theories. Indeed, the time-independent equation (3.15) has been obtained from the time-dependent 10D action (2.7) plus (2.22) (or equivalently (2.1) plus (2.17)). But the resulting 6D equation that we finally obtain is the time-independent equation that coincides with the usual 6D equation of the time-independent theories. This result is consistent with the statement in Sec. 3.2 that the 6D effective action is time-independent, and it defines a time-independent theory on the 6D internal sector. Since $\tilde{\mathcal F}_{(5)}$ is self-dual, the equation of motion (3.15) becomes the Bianchi identity as it is:
\begin{equation}
d \tilde{\mathcal F}_{(5)} = {\mathcal H}_{(3)} \wedge {\mathcal F}_{(3)} + 2 \kappa^2_{10} \mu_0 \rho_3^{\rm loc} \,\,,
\end{equation}
where we have used ${\mathcal G}_{(3)} = {\mathcal F}_{(3)} - i {\rm Im} \tau \,{\mathcal H}_{(3)}$. (3.16) is time-independent equation as sure as (3.15) is. Under integration, (3.16) gives the tadpole cancellation
\begin{equation}
\frac{1}{2\kappa^2_{10} T_0} \int_{\mathcal M} {\mathcal H}_{(3)} \wedge {\mathcal F}_{(3)} + Q_3^{\rm loc} =0\,\,,
\end{equation}
which is also precisely the same equation as (2.16) of the usual time-independent theory.

So far, we have seen that the final 6D equations that appear to the 6D observer are the time-independent equations like (3.15) and (3.16), which are entirely given in terms of the time-independent $n$-form fields $\tilde{\mathcal F}_{(5)}$, ${\mathcal H}_{(3)}$ and ${\mathcal F}_{(3)}$. Since the 6D equations do not contain $\tilde{F}_{(5)}$, $H_{(3)}$ and $F_{(3)}$, these time-dependent $n$-form fields are irrelevant to an observer sitting in the six-dimensional internal space. Indeed, the 3-form fields that are associated with the geometry and topology of compactified internal dimensions are these ${\mathcal F}_{(3)}$ and ${\mathcal H}_{(3)}$. To the 6D observer the $n$-form fields $\tilde{F}_{(5)}$, $H_{(3)}$ and $F_{(3)}$ are invisible objects because they never appear in the 6D equations when the moduli and scale factors all evolve in time at the same rate in such a way that they satisfy the constraint equations (3.3) and (3.4). Since the whole resulting equations (and $n$-form fields) of this section precisely coincide with those equations (and $n$-form fields) of the usual time-independent theory, we find that the inconsistencies like those discussed at the end of Sec. 2.1 do not exist anymore, and in the following discussions in Sec. 3 we will follow the well-known procedures used in the existing theories.

\vskip 0.3cm
\hspace{-0.6cm}{\bf \large 3.4 Scalar potential}
\vskip 0.15cm

Now we define the superpotential as
\begin{equation}
{\mathcal W} = \int {\mathcal G}_{(3)} \wedge \Omega \,\,,
\end{equation}
where $\Omega$ is the holomorphic (3,0)-form of the CY three-fold. This ${\mathcal W}$ has no explicit time-dependence coming from $\eta (\hat t)$. But it depends on the complex structure moduli $z^{\alpha}$ (and the dilaton modulus $\tau$) because $\Omega$ (and ${\mathcal G}_{(3)}$) depends on $z^{\alpha}$ (and $\tau$), respectively. The flux induced potential for the complex structure moduli can be obtained from the ${\mathcal G}_{mnp}^{+} {\bar {\mathcal G}}^{+mnp}$ term in (3.8). In the Einstein frame this potential takes the form (see (A.12) in Appendix)
\begin{equation}
{\mathcal V}_{\rm scalar} =  - \frac{1}{2 \kappa_{10}^2}  \frac{1}{{\rm Im} \tau}  \frac{\eta^2}{\beta^{12}_{\rm E}} \int_{\mathcal M} {\mathcal G}_{(3)}^{+} \wedge \ast_{6} \, {\bar {\mathcal G}}_{(3)}^{+} \,\,,
\end{equation}
where we have used the approximation of the constant warp factor ($e^{2A_{\rm E}} =1$) which is valid in the large-radius limit \cite{5}. In (3.19) we have an additional factor ${\rm Im} \tau$ in the denominator, which does not exist in the original Lagrangian (3.8). This is because, ${\mathcal V}_{\rm scalar}$ in (3.19) has been obtained from the Einstein frame, while (3.8) is the Lagrangian in string frame. Note that the $G_{(3)} \cdot \bar{ G}_{(3)}$ term in the string frame action (2.1) becomes multiplied by $e^{\Phi}$ when we go to the Einstein frame (see (A.1)).

${\mathcal V}_{\rm scalar}$ in (3.19) contains the internal metric $h_{mn} (y)$ through the six-dimensional Hodge star and it induces a potential for the deformations of $h_{mn} (y)$. In the CY compactifications there are two types of deformations of $h_{mn} (y)$ which preserve the CY condition ${\mathcal R}_{mn} =0$. They are $\delta h_{a \bar{b}}$ ((1,1)-type) and $\delta h_{\bar{a} \bar{b}}$ (or $\delta h_{a b}$) ((2,1) or (1,2)-type) in the complex coordinate system. Among these, the (2,1) (or (1,2))-type deformation $\delta h_{\bar{a} \bar{b}}$ (or $\delta h_{a b}$) is identified with the complex structure deformations by the equation \cite{8-1}
\begin{equation}
\delta h_{\bar{a} \bar{b}} = - \frac{1}{\| \Omega \|^2} \, {\bar{\Omega}}_{\bar{a}}^{~\,cd} (\chi_{\alpha})_{cd \,\bar{b}} \,\delta z^{\alpha} \,\,,~~~~~(\alpha = 1, \, \cdots \, h^{2,1} ) \,\,,
\end{equation}
where $\chi_{\alpha}$ is the holomorphic (2,1)-form. The above statement implies that the potential (3.19) can be used as a potential for the stabilization of the complex structure moduli.

Now we show that the potential (3.19) coincides with the standard $F$-term potential of the $N=1$, $D=4$ supergravity. Since ${\mathcal G}_{(3)}^{+}$ is IASD, $\ast_{6} {\mathcal G}_{(3)}^{+} =-i {\mathcal G}_{(3)}^{+}$, it can be decomposed as
\begin{equation}
{\mathcal G}_{(3)}^{+} = \alpha \Omega + {\bar{\beta}}^{\bar{\alpha}} {\bar{\chi}}_{\bar{\alpha}}\,\,,
\end{equation}
where ${\bar{\chi}}_{\bar{\alpha}}$ denotes the basis of $H^{(1,2)}$ and it is IASD as well as $\Omega$. From (3.21) one obtains
\begin{equation}
\alpha = \frac{ \int_{\mathcal M} {\mathcal G}_{(3)}^{+} \wedge {\bar{\Omega}}}{\int_{\mathcal M} \Omega \wedge {\bar{\Omega} }} = \frac{ \int_{\mathcal M} {\mathcal G}_{(3)} \wedge {\bar{\Omega}}}{ \int_{\mathcal M} \Omega \wedge {\bar{\Omega}} } \,\,, ~~~
{\bar{\beta}}^{\bar{\alpha}} = G^{\beta {\bar{\alpha}}} \frac{ \int_{\mathcal M} {\mathcal G}_{(3)}^{+} \wedge \chi_{\beta} }{ \int_{\mathcal M} \Omega \wedge {\bar{\Omega}}} = G^{\beta {\bar{\alpha}}} \frac{ \int_{\mathcal M} {\mathcal G}_{(3)} \wedge \chi_{\beta}}{\int_{\mathcal M} \Omega \wedge {\bar{\Omega}}} \,\,,
\end{equation}
where the metric $G_{\alpha {\bar{\beta}}}$ is defined by
\begin{equation}
G_{\alpha {\bar{\beta}}} = - \frac{\int_{\mathcal M} \chi_{\alpha} \wedge {\bar \chi}_{{\bar \beta}} }{\int_{\mathcal M} \Omega \wedge {\bar{\Omega} }} \,\,,~~~~~(G^{\alpha {\bar{\beta}}} =G^{-1}_{\alpha {\bar{\beta}}})\,\,.
\end{equation}
By (3.22), (3.21) now can be rewritten as
\begin{equation}
{\mathcal G}_{(3)}^{+} =  \frac{\Omega  \int {\mathcal G}_{(3)} \wedge {\bar{\Omega}}}{ \int \Omega \wedge {\bar{\Omega}} } + G^{\alpha {\bar{\beta}}} \, \chi_{\bar{\beta}} \, \frac{ \int {\mathcal G}_{(3)} \wedge \chi_{\alpha}}{\int \Omega \wedge {\bar{\Omega}}} \,\,.
\end{equation}
Substituting (3.24) into (3.19), and using $\ast_{6} {\bar{\Omega}}=i {\bar{\Omega}}$ and $\ast_{6} \chi_\alpha =i \chi_\alpha$ one finds that (3.19) becomes
\begin{equation}
{\mathcal V}_{\rm scalar} = i \frac{\eta^2}{\beta^{12}_{\rm E}}\, \frac{\int_{\mathcal M} {\mathcal G}_{(3)} \wedge {\bar{\Omega} \int_{\mathcal M} {\bar{\mathcal G}}_{(3)} \wedge {\Omega}} + G^{\alpha {\bar{\beta}}} \int_{\mathcal M} {\mathcal G}_{(3)} \wedge \chi_{\alpha} \int_{\mathcal M} {\bar{\mathcal G}}_{(3)} \wedge {\bar \chi}_{{\bar \beta}}}{{2 \kappa^2_{10} {\rm Im} \tau}\int_{\mathcal M} \Omega \wedge {\bar{\Omega} }}  \,\,.
\end{equation}

Now we introduce the covariant derivative ${\mathcal D}_I {\mathcal W} = \partial_I {\mathcal W} + (\partial_I {\mathcal K}) {\mathcal W}$, where ${\mathcal K}$ is the K$\ddot{\rm a}$hler potential. At the tree level, ${\mathcal K}$ is given by
\begin{equation}
{\mathcal K} = -3 \ln \big( -i (\rho - {\bar{\rho}}) \big) - \ln \big( -i (\tau - {\bar{\tau}}) \big) - \ln \Big( -i \int \Omega \wedge {\bar{\Omega}} \Big) \,\,,
\end{equation}
where $\rho$ is the K$\ddot{\rm a}$hler (radial) modulus. In the approximation of the constant warp factor the Einstein frame metric (A.8) can be written as
\begin{equation}
ds_{\rm E}^2 = e^{-6 u(x)} g_{\mu \nu} (x) dx^{\mu} dx^{\nu} + e^{2 u(x)} h_{mn} (y) dy^{m} dy^{n} \,\,,
\end{equation}
where $u(x)$ parameterizes the volume of the CY three-fold and in our case $e^{u(x)}$ is a rewrite of $\beta_{\rm E}(t)$, i.e. $e^{u(x)} = \beta_{\rm E}(t)$. Now the K$\ddot{\rm a}$hler modulus $\rho$ is defined by
\begin{equation}
\rho = \frac{b}{\sqrt{2}} +i e^{4u}  \,\,,
\end{equation}
where $b$ represents another axion which is dual to a two-form $a_{(2)}$, $da_{(2)} = e^{-8 u(x)} \ast_{4} db$, where $a_{(2)}$ is defined by the equation $A_{\mu\nu pq} = a_{\mu \nu} J_{pq}$ (where $J$ is the K$\ddot{\rm a}$hler form). In our discussion we consider the case of a single K$\ddot{\rm a}$hler modulus which characterizes the size of CY, i.e. the radial modulus $\rho$, for simplicity. Using (3.26) one finds that
\begin{equation}
{\mathcal D}_{\alpha} {\mathcal W} = \int_{\mathcal M} {\mathcal G}_{(3)} \wedge \chi_{\alpha} \,\,
\end{equation}
\begin{equation}
{\mathcal D}_{\tau} {\mathcal W} = -\frac{1}{(\tau - {\bar{\tau}})} \int_{\mathcal M} {\bar{\mathcal G}}_{(3)} \wedge \Omega \,\,,
\end{equation}
and
\begin{equation}
{\mathcal D}_{\rho} {\mathcal W} = \frac{-3}{(\rho - {\bar{\rho}})} {\mathcal W} \,\,.
\end{equation}
Substituting the above equations into (3.25) we finally obtain
\begin{equation}
{\mathcal V}_{\rm scalar} \sim \frac{1}{2 \kappa^2_{10}} \, \eta^2 (t) \,e^{\mathcal K} \Big( G^{a \bar{b}} {\mathcal D}_{a} {\mathcal W} \,\, {\overline{{\mathcal D}_{b} {\mathcal W}}} - 3 |{\mathcal W}|^2 \Big) \,\,,
\end{equation}
where $G_{a \bar{b}} = \partial_a \partial_{\bar{b}} {\mathcal K}$ and the indices $a$, $b$ are summed all over the complex structure moduli $z_{\alpha}$, dilaton $\tau$ and the K$\ddot{\rm a}$hler modulus $\rho$.

\vskip 0.3cm
\hspace{-0.6cm}{\bf \large 3.5 Moduli stabilization}
\vskip 0.15cm

${\mathcal V}_{\rm scalar}$ in (3.32) precisely coincides with the standard $F$-term scalar potential because $\eta(t) {\mathcal W}$ is just the $G_{(3)}$-induced superpotential $W$
\begin{equation}
W= \int G_{(3)} \wedge \Omega \,\,.
\end{equation}
Though (3.32) itself is the original $F$-term scalar potential, in the followings we will just put
\begin{equation}
{\mathcal V}_{\rm scalar} \sim \frac{1}{2 \kappa^2_{10}} \,e^{\mathcal K} \Big( G^{a \bar{b}} {\mathcal D}_{a} {\mathcal W} \,\, {\overline{{\mathcal D}_{b} {\mathcal W}}} - 3 |{\mathcal W}|^2 \Big) \,\,,
\end{equation}
because $\eta^2 (t)$ in (3.32) does not play any important role in the following discussions in Sec. 3.5. Since (3.34) is given by the same form as the typical $F$-term scalar potential of the usual flux compactifications, it seems to be okay even if we borrow the existing scenarios for the moduli stabilization. In this paper we will accept the existing scenarios for the complex structure moduli as it is. However, in the case of the K$\ddot{\rm a}$hler (and dilaton) modulus the scenario of this paper is entirely distinguished from the existing scenarios as we will see in Sec. VII.

Now we start a brief review on the moduli stabilization scenario. Since ${\mathcal W}$ is of no-scale type, $|{\mathcal D}_{\rho} {\mathcal W}|^2$ term in (3.34) cancels $3|{\mathcal W}|^2$ term (Note that ${\mathcal D}_{\rho} {\mathcal W} = {-3 {\mathcal W}}/{(\rho - {\bar{\rho}})}$ (see (3.31)) and $G^{\rho {\bar {\rho}}} = - (\rho - {\bar{\rho}})^2 /3$.) and as a result (3.34) reduces to
\begin{equation}
{\mathcal V}_{\rm scalar} \sim \frac{1}{2 \kappa^2_{10}} \,e^{\mathcal K} \Big( G^{\alpha \bar{\beta}} {\mathcal D}_{\alpha} {\mathcal W} \,\, {\overline{{\mathcal D}_{\beta} {\mathcal W}}} +  G^{\tau \bar{\tau}} {\mathcal D}_{\tau} {\mathcal W} \,\, {\overline{{\mathcal D}_{\tau} {\mathcal W}}} \Big) \,\,.
\end{equation}
(3.35) is now positive definite and the global minimum of this potential is characterized by the conditions
\begin{equation}
{\mathcal D}_{\alpha} {\mathcal W} = \int_{\mathcal M} {\mathcal G}_{(3)} \wedge \chi_{\alpha} =0 \,\,,
\end{equation}
\begin{equation}
{\mathcal D}_{\tau} {\mathcal W} = - \frac{1}{ (\tau - {\bar {\tau}}) } \int_{\mathcal M} {\bar{\mathcal G}}_{(3)} \wedge {\Omega} =0 \,\,.
\end{equation}
These conditions are also associated with the supersymmetry of the background vacua. Supersymmetry remains unbroken when the $F$-terms ${\mathcal D}_{a} {\mathcal W}$ all vanish, i.e. when ${\mathcal D}_{a} {\mathcal W}=0$ for all $a$. When ${\mathcal W}=0$, this condition is satisfied because ${\mathcal D}_{\rho} {\mathcal W}$ also vanishes by (3.31). In this case we have Minkowski vacua with unbroken supersymmetry. If ${\mathcal W} \neq 0$, we obtain more interesting solution for the background vacua. Since ${\mathcal W} \neq 0$ implies ${\mathcal D}_{\rho} {\mathcal W} \neq 0$, the supersymmetry is broken, but ${\mathcal V}_{\rm scalar}$ still vanishes ($\lambda =0$) at the minimum of the potential. This result is interesting in the viewpoint of the cosmological constant problem. However, this cannot be the solution of the cosmological constant problem because the above result is restricted only to the classical level.

Turning back to the moduli stabilization problem, the $h^{2,1}$ complex structure moduli of the background CY are stabilized by the $h^{2,1}$ complex equations ${\mathcal D}_{\alpha} {\mathcal W} =0$ in (3.36). Perhaps the most interesting and well-known example of this scenario may be the one that given in \cite{5}. Though the background geometry of our internal space is basically a compact CY, the authors in \cite{5} considered a deformed conifold as our background manifold. Let us assume that there are $M$ units of ${\mathcal F}_{(3)}$ flux through an $A$-cycle and $-K$ units of ${\mathcal H}_{(3)}$ flux through a $B$-cycle which is dual to $A$,
\begin{equation}
\frac{1}{2 \pi \alpha^{\prime}} \int_{A} {\mathcal F}_{(3)} = 2 \pi M ~~~~~~~{\rm and}~~~~~~~ \frac{1}{2 \pi \alpha^{\prime}} \int_{B} {\mathcal H}_{(3)} = - 2 \pi K \,\,.
\end{equation}
Then the superpotential becomes
\begin{equation}
{\mathcal W} = \int {\mathcal G}_{(3)} \wedge \Omega = (2 \pi)^2 \alpha^{\prime} \, \big( M \int_{B} \Omega - K \tau \int_{A} \Omega \big) \,\,,
\end{equation}
by the Poincar$\acute{\rm e}$ duality. Now suppose that $z$ is the modulus representing the size of $S_3$ in the deformed conifold, and let the corresponding three-cycle (i.e. $S_3$) which collapses as $z \rightarrow 0$ be the $A$-cycle. Then (3.39) becomes
\begin{equation}
{\mathcal W} = (2 \pi)^2 \alpha^{\prime} \, ( M {\mathcal G}(z) - K \tau z )\,\,,
\end{equation}
where $z$ is the period of $\Omega$ along the $A$-cycle,
\begin{equation}
z = \int_{A} \Omega \,\,,
\end{equation}
as mentioned above, while ${\mathcal G}(z)$ is some complicated function of $z$ given by
\begin{equation}
{\mathcal G}(z) = \int_{B} \Omega = \frac{z}{2 \pi i} \ln z + {\rm holomorphic}\,\,.
\end{equation}
Using these results one finds that the covariant derivative of the superpotential ${\mathcal W}$ can be written as
\begin{equation}
{\mathcal D}_{z} {\mathcal W} = (2 \pi)^2 \alpha^{\prime} \, \Big( \frac{M}{2 \pi i} \ln z - i \frac{K}{g_s} + \cdots \Big) \,\,,
\end{equation}
in the limit in which ${K}/{g_s}$ is large, ${K}/{g_s} \gg 1$. Thus the modulus $z$ satisfying the condition ${\mathcal D}_{z} {\mathcal W} =0$ for the global minimum of the potential becomes
\begin{equation}
z \sim \exp \big( - 2 \pi K / M g_s \big) \,\,.
\end{equation}

The dilaton modulus is also stabilized by the minimum condition ${\mathcal D}_{\tau} {\mathcal W} =0$,
\begin{equation}
0 = {\mathcal D}_{\tau} {\mathcal W} \propto - \frac{1}{ (\tau - {\bar {\tau}}) }  \big( -K z {\bar{\tau}} +M {\mathcal G}(z) \big)\,\,.
\end{equation}
However, (3.45) cannot be satisfied if we insist on ${K}/{g_s} \gg 1$ to obtain a large hierarchy in scales, because the first term $-K z {\bar{\tau}}$ is exponentially small, while the second term is of order one,
${\mathcal G}(0) = O(1)$, when $z \rightarrow 0$. Hence in this case the equation ${\mathcal D}_{\tau} {\mathcal W} =0$ becomes meaningless and the dilaton is just assumed to be frozon. This problem may be avoided if we turn on additional fluxes on the different cycles (see \cite{5} and references therein) but it makes the things complicated. In this paper we basically accept the existing scenarios for the stabilization of the complex structure moduli described so far. But in the case of the K$\ddot{\rm a}$hler (and dilaton) modulus the scenario of this paper is entirely distinguished from the traditional KKLT type models. We have no nonperturbative corrections of the superpotential, and we have no anti D3-branes to obtain deSitter vacua. In the scenario of this paper the superpotential remains of no-scale type and the K$\ddot{\rm a}$hler modulus is stabilized by a set of dynamical (plus constraint) equations defined on the 4D spacetime as we will see in Sec. VII.

\vskip 0.3cm
\hspace{-0.6cm}{\bf \large 3.6 Quantum violation of Bianchi identity}
\vskip 0.15cm

In Sec. 3.3 we obtain the Bianchi identity for the 5-form field $\tilde{\mathcal F}_{(5)}$. In this equation the time-independent 3-forms ${\mathcal F}_{(3)}$ and ${\mathcal H}_{(3)}$ are both closed and satisfy the Bianchi identities
\begin{equation}
d{\mathcal F}_{(3)} =0 ~~~~~ {\rm and} ~~~~~ d{\mathcal H}_{(3)}=0\,\,,
\end{equation}
of the classical level. But on the contrary, the time-dependent 3-form fields $F_{(3)}$ and $H_{(3)}$ do not satisfy these Bianchi identities because $dF_{(3)}$, for instance, becomes
\begin{equation}
dF_{(3)} = d \big( \eta(t) {\mathcal F}_{(3)} \big) = d \eta(t) \wedge {\mathcal F}_{(3)} \,\,,
\end{equation}
which do not vanish if $\eta(t)$ is not a constant. In this final subsection we will briefly check up the Bianchi identities for the time-dependent 3-forms $F_{(3)}$ and $H_{(3)}$ for the completeness of our discussion.

In Sec.7 it will be shown that the quantum corrections of the total action ${\hat S}_{\rm total}$ is closely related with the constant $c$ by the equation (7.49). In (7.49), $Q$ is defined by (7.13) where $\delta_Q {\hat S}_{\rm total}$ represents quantum corrections of ${\hat S}_{\rm total}$, while the constant $c$ on the right hand side denotes the nonzero values acquired from the string corrections in $\Delta V$ (see (7.30)). Equation (7.49) is quite unusual equation which does not appear in the conventional theories. (7.49) implies that in the absence of string corrections (i.e. when $c=0$) the quantum corrections $\delta_Q {\hat S}_{\rm total}$ in $Q$ are entirely suppressed to zero (i.e. $\delta_Q {\hat S}_{\rm total}=0$). In our self-tuning mechanism this happens in the supergravity approximation. In the low energy supergravity the constant $c$ vanishes because in that case the string corrections are absent, and therefore the sum of all quantum corrections in $\delta_Q {\hat S}_{\rm total}$ is suppressed to zero as mentioned above (Indeed, the quantum corrections in $\delta_Q {\hat S}_{\rm total}$ are all gauged away at the supergravity level (see Sec. 5.2).) and the theory effectively reduces to the classical level. Indeed, in this case the set of 4D equations for the moduli in Sec. 7.1 is solved by $\beta(t) = \gamma(t) = 1$ (see (7.28)) which then gives
\begin{equation}
\eta(t) = 1 \,\,
\end{equation}
from (3.6). (3.48) means that $F_{(3)}$ and $H_{(3)}$ are identical with ${\mathcal F}_{(3)}$ and ${\mathcal H}_{(3)}$ at the supergravity level. Therefore $F_{(3)}$ and $H_{(3)}$ satisfy the same Bianchi identity as (3.46)
\begin{equation}
dF_{(3)} =0 ~~~~~ {\rm and} ~~~~~ dH_{(3)}=0 \,\,.
\end{equation}

Now we turn to the case with nonzero string corrections. In this case, $c$ takes nonzero values and therefore the quantum correction $\delta_Q {\hat S}_{\rm total}$ revives by (7.49). Since $\delta_Q {\hat S}_{\rm total}$ takes nonzero values now, we expect that the Bianchi identities in (3.49) would be modified by the quantum effect. Indeed, in the case of nonzero string corrections the moduli $\beta(t)$ and $\gamma(t)$ satisfying the set of 4D equations in Sec. 7.2 are not the constants anymore. They are now functions of time $t$ as in (7.63). The moduli $\beta(t)$ in (7.63) can be written as $\beta(t) = 1+ \delta_Q \beta(t)$, where the modification $\delta_Q \beta(t)$ ($\cong 0$) reflects the quantum effects due to $\delta_Q {\hat S}_{\rm total}$ because $\beta(t)$ reduces back to $\beta(t) = 1$ in the supergravity limit where $\delta_Q {\hat S}_{\rm total}$ is suppressed to $\delta_Q {\hat S}_{\rm total} =0$. $\eta(t)$ also can be written as
\begin{equation}
\eta(t) = 1+ \delta_Q \eta(t) \,\,, ~~~(\delta_Q \eta(t) \cong 0) \,\,,
\end{equation}
because $\eta(t)$ is related with $\beta(t)$ and $\gamma(t)$ by the equation (3.6). In (3.50) the nonzero $\delta_Q \eta(t)$ arises from the quantum correction $\delta_Q {\hat S}_{\rm total}$ as in $\beta(t)$.

The analysis given so far shows that the modification of the Bianchi identity in (3.47) is entirely due to the quantum corrections in $\delta_Q {\hat S}_{\rm total}$. Classically, the Bianchi identity is satisfied identically (Eqs. (3.46) and (3.49)). But in the presence of quantum effects it is known that the tree level Bianchi identity can be violated as in (3.47). Such a modified Bianchi identity may have implications for the quantizations of the theory and consistency with the geometry and the topology of the internal manifolds. Hence any modification to the Bianchi identity must be carefully justified and it must be consistent with the other physical structures of the theory. However, in our scenario this quantum violation of the Bianchi identity only occurs at the stage of the full string theory. It does not occur at the supergravity level as we have seen so far. Moreover, the six-dimensional internal space is always time-independent world whose dynamics and geometrical structure are entirely determined by the time-independent equations (Sec. 3.3) regardless of whether we are at the level of supergravity or at the full string theory. After all, the 3-form fields associated with the geometry and topology of our compactified internal dimensions are ${\mathcal F}_{(3)}$ and ${\mathcal H}_{(3)}$ (not $F_{(3)}$ and $H_{(3)}$) and this 3-form fields satisfy the unmodified Bianchi identity (3.46).

\vskip 1cm
\hspace{-0.65cm}{\bf \Large IV. 4D effective action and cosmological constant}
\vskip 0.5cm
\setcounter{equation}{0}
\renewcommand{\theequation}{4.\arabic{equation}}

\vskip 0.3cm
\hspace{-0.6cm}{\bf \large 4.1 Total action $S_{\rm total}$}
\vskip 0.15cm

Coming back to the string frame we introduce the 4D effective action which is needed to define $\lambda$ (Sec. 4.3), and also needed to obtain 4D equations of motion for the scale factors $\beta_Q$, $\beta$ and $\mathcal A$ (Sec. VII). To find this action, we first consider the relations
\begin{equation}
\frac{\alpha^4  \beta^4}{\gamma^2} =\frac{\alpha^4}{\beta^4} \Gamma_R^2 \,\,,~~~ \frac{\alpha^2  \beta^6}{\gamma^2} =\frac{\alpha^2}{\beta^2}
\Gamma_R^2 \,\,,
\end{equation}
which can be obtained from (2.25) and (3.4). Using (4.1) (and also using (2.25) and (3.6)) one can rewrite $S_{\rm IIB}$ in (2.7) as
\begin{equation}
S_{\rm IIB} = \frac{1}{2 \kappa^2 } \int d^4 {\hat x} \sqrt{-{\hat g}_4}\, \frac{\alpha^2}{\beta^2} \, \Gamma_R^2 \, \hat{{\mathcal R}}_4^{(\rm eff)} ({\hat g}_{\mu\nu}, \alpha, \beta, \gamma) + \int d^4 {\hat x} \sqrt{-{\hat g}_4} \frac{\alpha^4}{\beta^4} \Gamma_R^2 \,({\hat S}_{\rm bulk} + {\hat
S}_{\rm topological})\,\,,
\end{equation}
where $2 \kappa^2 \equiv {2 \kappa_{10}^2 g_s^2}/ \big( \int d^6 y \sqrt{h_6}  \chi^{-1/2} \big)$ and ${\hat S}_{\rm bulk}$ is given by
\begin{equation}
{\hat S}_{\rm bulk} = \frac{1}{2 \kappa_{10}^2 g_s^2} \int d^6 y \sqrt{h_6} \Big({\mathcal R}_6 (h_{mn}) - {\mathcal L}_F  \Big)\,\,.
\end{equation}

Similarly, using (2.25) and (3.4) one can show that $S_{\rm brane}$ in (2.22) can be rewritten as
\begin{equation}
S_{\rm brane} = \int d^4 {\hat x} \sqrt{-{\hat g}_4}\, \frac{\alpha^4}{\beta^4} \Gamma_R^2 \, {\hat S}_{\rm brane} \,\,,
\end{equation}
where ${\hat S}_{\rm brane}$ is the brane action density
\begin{equation}
{\hat S}_{\rm brane} \equiv \int d^6 y \sqrt{h_6} \Big( - \frac{T_0}{g_s} \chi^{{1}/{2}}(y) L (\gamma ({\hat t}))+ \mu_0 \xi(y) \Big)
\delta^6 (y) \,\,,
\end{equation}
and in (4.5) $L(\gamma ({\hat t}\,))$ is defined by
\begin{equation}
L(\gamma ({\hat t}\,)) \equiv \lim_{y \rightarrow 0} \frac{\Gamma_{NS} (\Phi)}{\Gamma_{R} (\Phi)} = 1+ g_s e^{\Phi_S (0)} ({\hat T}_1 - {\hat
\mu}_1 ) \gamma ({\hat t}\,) + \cdots \,\,.
\end{equation}
$L(\gamma ({\hat t}\,))$ becomes $L(\gamma ({\hat t}\,)) =1$ when the branes are BPS (${\hat T}_n = {\hat \mu}_n$). In this case ${\hat
S}_{\rm brane}$ in (4.5) takes the tree level form
\begin{equation}
{\hat S}_{\rm brane} ({\rm tree}) = \int d^6 y \sqrt{h_6} \Big( - \frac{T_0}{g_s} \chi^{{1}/{2}}(y) + \mu_0 \xi(y) \Big)
\delta^6 (y) \,\,,
\end{equation}
and it can be shown that this ${\hat S}_{\rm brane} (\rm tree)$ vanishes by the field equations for $\chi(y)$ and $\xi(y)$ (see Ref. \cite{6} or Sec. 6.2 of this paper). Indeed the integrand of ${\hat S}_{\rm brane}$ acts as a D3-brane potential (see (2.24)) and it is known that it vanishes for $\mu_0 =T_0$ at the tree level. (We will show this in Sec. 6.2.) But once the brane supersymmetry is broken by the perturbations, ${\hat S}_{\rm brane}$ acquires nonvanishing correction terms coming from the quantum fluctuations and in this case ${\hat S}_{\rm brane}$ does not vanish
anymore.

Now the total action $S_{\rm total}$ can be obtained by adding (4.4) to (4.2). We have
\begin{equation}
S_{\rm total} = \frac{1}{2 \kappa^2 } \int d^4 {\hat x} \sqrt{-{\hat g}_4}\, \frac{\alpha^2}{\beta^2} \, \Gamma_R^2 {\hat{\mathcal R}}_4^{(\rm eff)} ({\hat g}_{\mu\nu}, \alpha, \beta, \gamma)
 + \int d^4 {\hat x} \sqrt{-{\hat g}_4} \frac{\alpha^4}{\beta^4} \, \Gamma_R^2 \, {\hat S}_{\rm total}
\,\,,
\end{equation}
where ${\hat S}_{\rm total}$ is defined by
\begin{equation}
{\hat S}_{\rm total} \equiv {\hat S}_{\rm bulk} + {\hat S}_{\rm brane}+{\hat S}_{\rm topological} \,\,,
\end{equation}
while $\hat{{\mathcal R}}_4^{(\rm eff)} ({\hat g}_{\mu\nu}, \alpha, \beta, \gamma)$ is given by (2.8) or (2.10). $S_{\rm total}$ in (4.8) will be identified as the 4D effective action in the next section, and from this action we will define $\lambda$ in Sec. 4.3, and obtain 4D equations of motion in Sec. VII.

\vskip 0.3cm
\hspace{-0.6cm}{\bf \large 4.2 $S_{\rm total}$ as a 4D effective action}
\vskip 0.15cm

(4.8) contains the curvature scalar of the 4D spacetime metric ${\hat g}_{\mu\nu}$ (see (2.8)) and it can be used as a 4D effective action containing gravity. However, the curvature term contained in (4.8) is not the standard Hilbert-Einstein action $\frac{1}{2 \kappa^2 } \int d^4 x \sqrt{-g_4}\,{\hat {\mathcal R}}_4$ of the gravity yet. So in order to obtain 4D action containing standard Hilbert-Einstein action we need some procedure given below.

In the first term of (4.8), $\hat{{\mathcal R}}_4^{(\rm eff)}$ was previously given by both (2.8) and (2.10). In this section we choose (2.10) to start our discussion. Using
\begin{equation}
\frac{\beta^3}{\gamma} = \frac{\Gamma_R}{\beta}\,\,,
\end{equation}
which follows from (2.25) and (3.4), one can rewrite (2.10) in more convenient form as
\begin{equation}
\sqrt{-{\hat g}_4}\, \frac{\alpha^2}{\beta^2} \,  \Gamma_R^2  \hat{{\mathcal R}}_4^{(\rm eff)} ({\hat g}_{\mu\nu}, \alpha, \beta, \gamma) = \frac{d}{d{\hat t}} \Bigg(\frac{{{\mathcal A}}^3 \beta_Q}{\alpha} \Bigg( 6 \Big(\frac{\dot{{\mathcal A}}}{{\mathcal A}} + \frac{\dot{\beta}_Q}{\beta_Q} \Big)
+12 \frac{\dot{\beta}}{\beta} \Bigg) \Bigg) ~~~~~~~~~~~~~~~~~~\nonumber
\end{equation}
\begin{equation}
~~~~~~~~~~~~~~~~~~~~~~~+ \frac{{{\mathcal A}}^3 \beta_Q}{\alpha}  \Bigg(- 6 \Big(\frac{\dot{{\mathcal A}}}{{\mathcal A}} \Big)^2 + 2
\Big(\frac{\dot{\beta}_Q}{\beta_Q} \Big)^2 + 6 \Big(\frac{\dot{\beta}}{\beta}\Big)^2 \Bigg) \,\,,
\end{equation}
where ${\mathcal A}$ and $\beta_Q$ are defined, respectively, by
\begin{equation}
{\mathcal A} \equiv a \alpha \frac{\Gamma_R}{\beta}\,\,,~~~~~\beta_Q \equiv \frac{\beta}{\Gamma_R} \,\,.
\end{equation}
Now we make a coordinate transformation ${\hat t} \rightarrow t$ defined by
\begin{equation}
dt = \frac{\alpha}{\beta_Q}  d {\hat t} \,\,.
\end{equation}
Then, using (4.11) (and also (4.13)) one can rewrite the first term of (4.8) as
\begin{equation}
\frac{1}{2 \kappa^2} \int d^4 {\hat x} \sqrt{-{\hat g}_4}\, \frac{\alpha^2}{\beta^2} \, \Gamma_R^2 {{\hat{\mathcal R}}}_4^{(\rm eff)} ({\hat g}_{\mu\nu}, \alpha, \beta, \gamma) = \frac{1}{2 \kappa^2} \int d^3 {\vec x} \int dt \Bigg[ \frac{d}{dt} \Bigg(
{\mathcal A}^3 \Bigg( 6 \Big(\frac{\dot{{\mathcal A}}}{{\mathcal A}} + \frac{\dot{\beta}_Q}{\beta_Q} \Big) +12 \frac{\dot{\beta}}{\beta} \Bigg) \Bigg) \nonumber
\end{equation}
\begin{equation}
+{\mathcal A}^3 \Bigg(- 6 \Big(\frac{\dot{{\mathcal A}}}{{\mathcal A}} \Big)^2 + 2 \Big(\frac{\dot{\beta}_Q}{\beta_Q} \Big)^2 + 6
\Big(\frac{\dot{\beta}}{\beta}\Big)^2 \Bigg) \Bigg]\,\,,
\end{equation}
where the "dot" now denotes the derivative with respect to $t$.

(4.14) is the 4D effective action for the curvature defined on the 4D sector $(t, {\vec x})$ of the 10D spacetime whose metric is now given by
(see (2.2), (4.12), (4.13) and (4.10))
\begin{equation}
ds_{10}^2 = \frac{\gamma^2 (t)}{\beta^6 (t)}  e^{A(y)} g_{\mu \nu} (x) dx^{\mu} dx^{\nu} + \beta^2 (t) e^{\Phi_S (y) -A(y)} h_{mn} (y) dy^m dy^n \,\,,
\end{equation}
where the 4D metric $g_{\mu \nu} (x) dx^{\mu} dx^{\nu}$ is defined by
\begin{equation}
g_{\mu \nu} (x) dx^{\mu} dx^{\nu}= -dt^2 + {\mathcal A}^2 (t) d{\vec x}_3^2 \,\,.
\end{equation}
Indeed (4.14) can be recast into
\begin{equation}
\frac{1}{2 \kappa^2 } \int d^4 {\hat x} \sqrt{-{\hat g}_4}\, \frac{\alpha^2}{\beta^2} \, \Gamma_R^2 {\hat{\mathcal R}}_4^{(\rm eff)} ({\hat g}_{\mu\nu}, \alpha, \beta, \gamma) = \frac{1}{2 \kappa^2} \int d^3 {\vec x} \int dt \sqrt{-g_4} {{\mathcal R}}_4^{(\rm eff)}  (g_{\mu\nu}, \beta_Q, \beta)
\end{equation}
with ${{\mathcal R}}_4^{(\rm eff)}  (g_{\mu\nu}, \beta_Q, \beta)$ defined by
\begin{equation}
{{\mathcal R}}_4^{(\rm eff)}  (g_{\mu\nu}, \beta_Q, \beta) = {{\mathcal R}}_4 (g_{\mu\nu})+6\frac{d}{dt} \Big(\frac{\dot{\beta}_Q}{\beta_Q} \Big)
+12\frac{d}{dt}\Big(\frac{\dot{\beta}}{\beta} \Big) +2\Big(\frac{\dot{\beta}_Q}{\beta_Q} \Big)^2 + 6 \Big(\frac{\dot{\beta}}{\beta} \Big)^2 +18
\frac{\dot{{\mathcal A}}}{{\mathcal A}} \frac{\dot{\beta}_Q}{\beta_Q} +36 \frac{\dot{{\mathcal A}}}{{\mathcal A}} \frac{\dot{\beta}}{\beta}
\,\,,
\end{equation}
where ${{\mathcal R}}_4 (g_{\mu\nu})/\sqrt{-g_4}$ are the Ricci-Scalar/determinant of the 4D metric (4.16), respectively. So we have ${{\mathcal
R}}_4 (g_{\mu\nu}) = 6\Big(\frac{\ddot{{\mathcal A}}}{{\mathcal A}} + \big(\frac{\dot{{\mathcal A}}}{{\mathcal A}}\big)^2 \Big)$ and
$\sqrt{-g_4}= {\mathcal A}^3$. ${{\mathcal R}}_4^{(\rm eff)} (g_{\mu\nu}, \beta_Q, \beta)$ in (4.18) reduces to ${{\mathcal R}}_4 (g_{\mu\nu})$ in
the time-independent limit $\dot \beta_Q =\dot \beta =0$. So in the coordinate system where the metric is given by (4.15), (4.17) becomes the
standard 4D Hilbert-Einstein action of the metric (4.16) in the limit $\dot \beta_Q =\dot \beta =0$. Now using (4.17) (and also using
(4.12) and (4.13)) one can show that (4.8) finally takes the form
\begin{equation}
S_{\rm total} =  \frac{1}{2 \kappa^2 }  \int d^3 {\vec x} \int dt \sqrt{-g_4} {{\mathcal R}}_4^{(\rm eff)}  (g_{\mu\nu}, \beta_{Q}, \beta) +  \int
d^3 {\vec x} \int dt \sqrt{-g_4} \, \frac{\beta_Q^2}{\beta^2} {\hat S}_{\rm total} \,\,.
\end{equation}

\vskip 0.3cm
\hspace{-0.6cm}{\bf \large 4.3 4D cosmological constant}
\vskip 0.15cm

$S_{\rm total}$ in (4.19) can be rewritten as
\begin{equation}
S_{\rm total}= \frac{1}{2 \kappa^2 } \int d^3 {\vec x} \int dt \sqrt{-g_4} \Big( {{\mathcal R}}_4 (g_{\mu\nu}) - 2 \lambda \Big) \,\,,
\end{equation}
where $\lambda$ is the cosmological constant defined by
\begin{equation}
\lambda = -\kappa^2 \Big( \frac{\beta_Q}{\beta} \Big)^2 {\hat S}_{\rm total} - \frac{1}{2} \Delta {{\mathcal R}}_4 (g_{\mu\nu}, \beta_Q, \beta)
\,\,,
\end{equation}
where $\Delta {{\mathcal R}}_4 (g_{\mu\nu}, \beta_Q, \beta)$ represents the extra terms in (4.18):
\begin{equation}
\Delta {{\mathcal R}}_4 (g_{\mu\nu}, \beta_Q, \beta) \equiv 6\frac{d}{dt} \Big(\frac{\dot{\beta}_Q}{\beta_Q} \Big)
+12\frac{d}{dt}\Big(\frac{\dot{\beta}}{\beta} \Big) +2\Big(\frac{\dot{\beta}_Q}{\beta_Q} \Big)^2 + 6 \Big(\frac{\dot{\beta}}{\beta} \Big)^2 +18
\frac{\dot{{\mathcal A}}}{{\mathcal A}} \frac{\dot{\beta}_Q}{\beta_Q} +36 \frac{\dot{{\mathcal A}}}{{\mathcal A}} \frac{\dot{\beta}}{\beta}
\,\,.
\end{equation}
In (4.21), $\beta$ and $\beta_Q$ (and also $\dot \beta$ and $\dot \beta_Q$) all represent their present values because (4.21) is the cosmological constant of the present universe. (4.21) suggests that not only are the
(quantum corrections of) ${\hat S}_{\rm total}$ the contributions to $\lambda$. The time evolution (i.e. nonzero $\dot \beta_Q$ and $\dot
\beta$) of the internal dimensions also contributes to $\lambda$ in the scenario of this paper. But in the limit $\beta_Q (t) = \beta (t) =1$, (4.21) reduces to
the equation
\begin{equation}
\lambda = - \kappa^2 {\hat S}_{\rm total} \,\,,
\end{equation}
which coincides with Eq. (3.15) of Ref. \cite{3} if we ignore ${\hat S}_{\rm topological}$ in ${\hat S}_{\rm total}$.\footnote{In (3.15) of Ref. \cite{3} the topological term ${\hat S}_{\rm topological}$ has been omitted for convenience. But this omission of ${\hat S}_{\rm topological}$ in \cite{3} will not change the story of Ref. \cite{3} at all because this ${\hat S}_{\rm topological}$ is always gauged away together with ${\hat S}_{\rm bulk}$ and ${\hat S}_{\rm brane}$ in ${\hat S}_{\rm total}$ anyway. (See Sec. 1 of Ref. \cite{3} or see Sec. 5.2 of this paper.)}

Going back to Sec. 3.2 we see that ${\mathcal L}_F$ in (3.8) includes both kinetic and potential terms, where the former takes the form $K=h^{mn}K_{mn}$ with $K_{mn}$ given by $K_{mn} = \sum_{A,B} F_{AB}(\phi_C) \partial_m \phi_A \partial_n \phi_B$, where $\phi_A$ represent the 6D scalars such as $\Phi_S$, $A$ and $\xi$ etc. So in the kinetic part of ${\mathcal L}_F$, $K_{mn}$ does not involve any metric $h^{mn}$, but the potential part $V$ of ${\mathcal L}_F$ (i.e. ${\mathcal G}_{mnp}^+ {\bar {\mathcal G}}^{+mnp}$ term in the case of (3.8)) includes $h^{mn}$: $V=V(h^{mn}, \,\,{\mathcal G}_{mnp}^+ , \,\,\cdots )$, where $V$ is related to the string frame scalar potential ${\mathcal V}_{\rm scalar}^S$ of the $N=1$ supergravity by the equation
\begin{equation}
{\mathcal V}_{\rm scalar}^S = - \frac{1}{2 \kappa^2_{10} g_s^2} \int d^6 y \sqrt{h_6} \, V \,\,,
\end{equation}
(see also Eq.(5.20)). In any case, ${\mathcal L}_F$ in (3.8) generally takes the form
\begin{equation}
{\mathcal L}_F = K - V \,\,,~~~~~(K=h^{mn}K_{mn})\,\,,
\end{equation}
and if we substitute (3.12) (with ${\mathcal L}_F$ given by (4.25)) into (3.11), we obtain (after contracting the indices $m$ and $n$)
\begin{equation}
{\mathcal R}_6 -{\mathcal L}_F - \frac{1}{2} ({\mathcal N}-1)V + \frac{3}{2} c \chi^{-1/2} =0\,\,,
\end{equation}
where ${\mathcal N}$ is defined by ${\mathcal N} \equiv h^{mn} \frac{\partial}{\partial h^{mn}}$ and $c$ (which was defined by (3.9)) is now
given by
\begin{equation}
c = \frac{\int d^3 {\vec x} \int dt \sqrt{-g_4} {{\mathcal R}}_4^{(\rm eff)}  (g_{\mu\nu}, \beta_Q, \beta)}{\int d^3 {\vec x} \int dt \sqrt{-g_4}
\frac{\beta_Q^2}{\beta^2}} \,\,.
\end{equation}
(See (4.1) and (4.17) together with (4.12) and (4.13).)

(4.27) can be rewritten as
\begin{equation}
c = \Big(\frac{\beta}{\beta_Q}\Big)^2 {{\mathcal R}}_4^{(\rm eff)} (g_{\mu\nu}, \beta_Q, \beta) \,\,,
\end{equation}
because  in (4.27) $g_{\mu\nu}$, $\beta$ and $\beta_Q$ (and their derivatives) all represent their present values of our universe and therefore ${{\mathcal R}}_4^{(\rm eff)} (g_{\mu\nu}, \beta_Q, \beta)$ and $\frac{\beta_Q^2}{\beta^2}$ in (4.27) are virtually constants in a short time interval of the present stage of our universe. Indeed, ${{\mathcal R}}_4 (g_{\mu\nu})$ in ${{\mathcal R}}_4^{(\rm eff)} (g_{\mu\nu}, \beta_Q, \beta)$ is generally given by ${{\mathcal R}}_4 (g_{\mu\nu}) = 4 \lambda$ for the maximally symmetric spacetime (which means that we put ${\mathcal A}(t) = e^{\sqrt{\frac{\lambda}{3}}t}$ in the metric (4.16)) and similarly $\frac{\dot{{\mathcal A}}}{{\mathcal A}}$,
$\frac{\ddot{{\mathcal A}}}{{\mathcal A}}$, $\frac{\dot{\beta}_Q}{\beta_Q}$ and $\frac{\dot{\beta}}{\beta}$ will be essentially taken as constants in the 4D equations of motion in Sec.VII. So (4.28) becomes
\begin{equation}
c =\Big(\frac{\beta}{\beta_Q}\Big)^2 \Big( 4 \lambda + \Delta {{\mathcal R}}_4 \Big) \,\,
\end{equation}
for the metric (4.16) with ${\mathcal A}(t)$ given by ${\mathcal A}(t) = e^{\sqrt{\frac{\lambda}{3}}t}$.

Now we integrate (4.26) and use (4.3) and (4.29) to obtain
\begin{equation}
{\hat S}_{\rm bulk} = - \frac{3}{4 \kappa^2} \Big(\frac{\beta}{\beta_Q}\Big)^2 \Big( 4 \lambda + \Delta {{\mathcal R}}_4 \Big) + \frac{1}{4 \kappa^2_{10} g_s^2} \int d^6 y \sqrt{h_6} ({\mathcal N}-1) V \,\,.
\end{equation}
Also, substituting (4.30) into (4.21) we finally get
\begin{equation}
\lambda = \Big(\frac{\beta}{\beta_Q}\Big)^2 \Big( \frac{\kappa^2}{8 \kappa^2_{10} g_s^2} \int d^6 y \sqrt{h_6} ({\mathcal N}-1) V + \frac{\kappa^2}{2} ({\hat S}_{\rm brane} +{\hat S}_{\rm topological}) \Big) - \frac{1}{8} \Delta {{\mathcal R}}_4 \,\,,
\end{equation}
which is the generalization of Eq.(3.20) of Ref.\cite{3}.\footnote{Again, Eq.(3.20) of Ref. \cite{3} must include ${\hat S}_{\rm topological}$. That is, the term $\frac{\kappa^2}{2} {\hat S}_{\rm brane}$ must be replaced by $\frac{\kappa^2}{2} ({\hat S}_{\rm brane} +{\hat S}_{\rm topological})$ in that equation.} Note that we have not set $ {\mathcal R}_{mn} =  {\mathcal R}_6 =0$ in the whole
procedure of obtaining (4.31) because these quantities can take nonzero values off-shell in general. But $\lambda$ in (4.31) does not include ${\mathcal R}_6 (h_{mn})$ or ${\mathcal R}_{mn} (h_{mn})$. They cancelled out during the process of obtaining (4.31) and as a result $\lambda$ becomes independent of ${\mathcal R}_6 (h_{mn})$.

\vskip 1cm
\hspace{-0.65cm}{\bf \Large V. Self-tuning mechanism}
\vskip 0.5cm
\setcounter{equation}{0}
\renewcommand{\theequation}{5.\arabic{equation}}

\vskip 0.3cm
\hspace{-0.6cm}{\bf \large 5.1 Self-tuning equation for $\lambda$}
\vskip 0.15cm

In this section we use the self-tuning mechanism in \cite{3} to obtain a self-tuning equation for $\lambda$ of our present model in
which the internal dimensions are allowed to evolve with time. We first substitute ${\mathcal L}_F$ in (4.26) into (3.12) to get
\begin{equation}
T_{mn} = 2 ({\mathcal R}_{mn} - \frac{1}{2} h_{mn} {\mathcal R}_{6}) + \frac{1}{2} h_{mn} ({\mathcal N} - 1) V - \frac{\partial}{\partial
h^{mn}}({\mathcal N} - 1) V  - \frac{3}{2} c \chi^{-1/2} h_{mn} \,\,.
\end{equation}
Next, substitute (5.1) into (3.11) and contract $m$ and $n$. Then we obtain a constraint equation
\begin{equation}
c = - \frac{1}{3} \chi^{1/2} ({\mathcal N} - 1)({\mathcal N} -3) V \,\,,
\end{equation}
which will be identified with the self-tuning equation for $\lambda$ as we will see soon. (See also Sec. 5.3 for the modified version of (5.2).) (5.2) demands that $c$ must vanish if the potential density $V$ has a certain tensor structure described below.

The constraint equation (5.2) can be generalized by the procedure presented in \cite{3}. The most general form of the constraint equation
for $c$ is
\begin{equation}
c =\frac{1}{6} \chi^{1/2} ({\mathcal N} -1) ({\mathcal N} -3) (1-3 b_0 \Pi ({\mathcal N})) V \,\,,
\end{equation}
where $b_0$ is a constant and $\Pi ({\mathcal N})$ is an operator of the form
\begin{equation}
\Pi ({\mathcal N}) = \sum_{k} c_k ({\mathcal N} -n_1 ) \cdots ({\mathcal N} -n_k ) \,\,,
\end{equation}
where $n_k$ are integers. Now (5.2) (or equivalently (5.3)) requires that $c$ must vanish at least if $V$ belongs to $V_n$ ($V \in V_n$) with $n=1$ or 3, where $V_n$
represents a class of potential densities satisfying
\begin{equation}
{\mathcal N} V_n = n V_n \,\,.
\end{equation}
In the usual flux compactifications where the three-form fluxes are used to stabilize the moduli fileds, $V$ basically belongs to $V_3$ and therefore $c$ must vanish by (5.3), i.e. we have $c=0$.

In the present paper we consider the type IIB compactifications where the complex structure moduli are stabilized by the three-form flux ${\mathcal G}_{(3)}$. Hence at the supergravity level the potential density is given by $V \propto {\mathcal G}_{mnp}^+ {\bar {\mathcal G}}^{+mnp}$ (i.e. $V \in V_3$) and therefore $c$ vanishes by (5.2) (or (5.3)) because $({\mathcal N} -3) V_3 =0$. Now from (4.29) $\lambda$ becomes
\begin{equation}
c=0 ~~~\rightarrow~~~ \lambda = - \frac{1}{4} \Delta {\mathcal R}_4 \,\,,
\end{equation}
where $\Delta {\mathcal R}_4$ is given by (4.22). In Sec. 7.1 we will show that $\beta(t)$ and $\beta_Q (t)$ become $\beta(t) = \beta_Q (t) =1$ in the case of the low-energy supergravity. In this case $\lambda$ in (5.6) reduces to $\lambda =0$ because $\Delta {\mathcal R}_4$ vanishes for $\beta(t) = \beta_Q (t) =1$. This shows that the constraint $c=0$ is equivalent to the fine-tuning $\lambda =0$ in the case of the supergravity approximation.

\vskip 0.3cm
\hspace{-0.6cm}{\bf \large 5.2  Constituents of $\lambda$}
\vskip 0.15cm

Now we have two independent equations for $\lambda$. We have (4.31), and also we have (5.6) which is equivalent to $c =0$. Since (5.6) follows
from the constraint $c =0$, it acts as a constraint (self-tuning) equation for $\lambda$. (We have just seen that (5.6) becomes the self-tuning $\lambda =0$ in the case of the supergravity approximation.)

The other equation (4.31), on the other hand, tells about the constituents of $\lambda$. For the potential density $V \in V_n$, it reduces to
\begin{equation}
\lambda = \Big(\frac{\beta_Q}{\beta}\Big)^2 \Big(- \frac{(n-1)}{4} \kappa^2 {\mathcal V}_{\rm scalar}^S + \frac{\kappa^2}{2} ({\hat S}_{\rm brane} +{\hat
S}_{\rm topological})\Big) - \frac{1}{8} \Delta {\mathcal R}_4   \,\,,
\end{equation}
which shows that $\lambda$ is composed of three parts: i.e. a scalar potential ${\mathcal V}_{\rm scalar}^S$, brane plus topological action
density ${\hat S}_{\rm brane}+{\hat S}_{\rm topological}$, and finally $\Delta {\mathcal R}_4$ coming from the
nonvanishing ${\dot{\beta}_Q}$ and $\dot{\beta}$. For $n=3$, (5.7) becomes
\begin{equation}
\lambda =  \frac{\kappa^2}{2} \Big(\frac{\beta_Q}{\beta}\Big)^2 \Big(-{\mathcal V}_{\rm scalar}^S + {\hat S}_{\rm brane} + {\hat S}_{\rm topological}
\Big) - \frac{1}{8} \Delta {\mathcal R}_4 \,\,,
\end{equation}
and this reduces to Eq. (3.44) of Ref. \cite{3} in the limit $\beta(t) = \beta_Q (t) =1$.\footnote{As in (3.15) and (3.20), (3.44) of Ref. \cite{3} must also contain ${\hat S}_{\rm topological}$ (See footnotes 3 and 4). Namely $\lambda$ must appear in the form $\lambda = \frac{\kappa^2}{2}  \big(-{\mathcal V}_{\rm scalar}^S + {\hat S}_{\rm brane} +{\hat S}_{\rm topological}\big)$ in the Eq.(3.44) of Ref. \cite{3}. But again, the omission of ${\hat S}_{\rm topological}$ will not change the result of Ref. \cite{3} at all because this ${\hat S}_{\rm topological}$ is always gauged away (together with ${\mathcal V}_{\rm scalar}$ and ${\hat S}_{\rm brane}$) by ${\mathcal E}_{\rm SB}$ as described in Ref. \cite{3} or Sec. 5.2 of this paper.}

Among the constituents in (5.8), ${\hat S}_{\rm brane}$ can be decomposed further into three parts. We have
\begin{equation}
{\hat S}_{\rm brane} = \Big( {\hat S}_{\rm brane}^{(NS)} ({\rm tree}) + {\hat S}_{\rm brane}^{(R)} ({\rm tree}) \Big) + \Big( \delta_Q {\hat
S}_{\rm brane}^{(NS)} + \delta_Q {\hat S}_{\rm brane}^{(R)}  \Big) - {\mathcal E}_{\rm SB} \,\,.
\end{equation}
(See Sec. 3.5 of Ref. \cite{3} for this.) In (5.9), ${\hat S}_{\rm brane}^{(NS)} ({\rm tree})$ and ${\hat S}_{\rm brane}^{(R)} ({\rm tree})$ are
the (NS-NS and R-R parts of the) tree-level actions and they always cancel out by field equations for the BPS D3-branes. (We will show this
briefly in Sec. 6.2.) The next terms $\delta_Q {\hat S}_{\rm brane}^{(NS)}$ and $\delta_Q {\hat S}_{\rm brane}^{(R)}$ arise from $\rho_{\rm
vac}$ and $\delta \mu$, respectively, and they represent quantum fluctuations of the gravitational and standard model degrees of freedom with support on the
D3-branes. So $\delta_Q {\hat S}_{\rm brane}^{(NS)} + \delta_Q {\hat S}_{\rm brane}^{(R)}$ corresponds to the gravitational plus electroweak
and QCD vacuum energies of the standard model configurations of the brane region.

The last term ${\mathcal E}_{\rm SB}$ is a supersymmery breaking term, which originates from a gauge symmetry breaking of $A_{(4)}$ arising at the quantum level in the brane region. (See Secs. 3.4 and 3.5 of Ref. \cite{3} for the details.) ${\mathcal E}_{\rm SB}$ takes
the form
\begin{equation}
{\mathcal E}_{\rm SB} = - \int d^6 y \sqrt{h_6} \delta \mu_{\rm T}^m (\Phi) f_m (y) \delta^6 (y) \,\,,
\end{equation}
where $\delta \mu_{\rm T}^m$ represent quantum excitations on the brane with components along the transverse directions of the D3-branes and
$f_m (y)$ are arbitrary functions of $y^m$ representing (derivatives of) local gauge parameters. Since ${\mathcal E}_{\rm SB}$ contains the
gauge parameters $f_m (y)$, it has its own gauge arbitrariness.

Among the five terms in (5.9), ${\hat S}_{\rm brane}^{(NS)} ({\rm tree}) + {\hat S}_{\rm brane}^{(R)} ({\rm tree})$ vanishes for the BPS branes as mentioned above. But the quantum fluctuations $\delta_Q {\hat S}_{\rm brane}^{(NS)} + \delta_Q {\hat S}_{\rm brane}^{(R)}$ on the branes do not vanish when supersymmetry of the brane region is broken. So at the quantum level, ${\hat S}_{\rm brane}$ contained in (5.8) acquires nonzero contributions from these $\delta_Q {\hat S}_{\rm brane}^{(NS)} + \delta_Q {\hat S}_{\rm brane}^{(R)}$, and ${\mathcal E}_{\rm SB}$ in (5.10). Similarly, ${\hat S}_{\rm topological}$ also acquires nonzero contributions only from the quantum corrections $\delta_Q {\hat S}_{\rm topological}$ as we will see in Sec. 6.2. ${\hat S}_{\rm topological}$ is proportional to the D3-brane potential $\Phi_{-} (y)$ (see (6.11)) and hence it vanishes at the ISD (tree level) background: i.e. we have ${\hat S}_{\rm topological} (\rm tree)=0$. After all this, $\lambda$ in (5.8) (see also (5.22)) finally reduces to
\begin{equation}
\lambda = \frac{\kappa^2}{2} \Big(\frac{\beta_Q}{\beta}\Big)^2 \delta_Q {\hat S}_{\rm total} - \frac{1}{8} \Delta {\mathcal R}_4   \,\,,
\end{equation}
where $\delta_Q {\hat S}_{\rm total}$ represents
\begin{equation}
\delta_Q {\hat S}_{\rm total} = -{\mathcal V}_{\rm scalar}^S + \delta_Q {\hat S}_{\rm brane}^{(NS)}+ \delta_Q {\hat S}_{\rm brane}^{(R)} + \delta_Q {\hat S}_{\rm topological} - {\mathcal E}_{\rm SB} \,\,.
\end{equation}
In (5.11), $\delta_Q {\hat S}_{\rm total}$ contains ${\mathcal E}_{\rm SB}$ which possesses its own gauge arbitrariness. So the fluctuations $\delta_Q {\hat S}_{\rm brane}^{(NS)} + \delta_Q {\hat S}_{\rm brane}^{(R)}$, $\delta_Q {\hat S}_{\rm topological}$, and also the remaining terms $-{\mathcal V}_{\rm scalar}^S$ and $-\frac{1}{8} \Delta {\mathcal R}_4$ in $\lambda$ are all gauged away (compensated) by ${\mathcal E}_{\rm SB}$ so that $\lambda$ in (5.11) always satisfies the fine-tuning equations found in Sec. VII.

\vskip 0.3cm
\hspace{-0.6cm}{\bf \large 5.3  Generalized equations in the case of nonzero ${\alpha^{\prime}}$-corrections}
\vskip 0.15cm

The self-tuning equation (5.2) has been obtained on an assumption that the potential density $V$ does not include any derivatives of the metric $h_{mn}$. Once $V$ includes such derivative terms the self-tuning equation (5.2) needs to be modified. The ${\alpha^{\prime}}$-corrections of the string theory usually include higher derivative terms made of Riemann and Ricci tensors. So if $V$ includes these terms, the self-tuning equations in Sec. 5.1 must be replaced by the modified equations because the Riemann and Ricci tensors consist of the first and second derivatives of $h_{mn}$. In this section we briefly derive the modified version of (5.2), which will be used in Sec. 7.2.

We start from the energy-momentum tensor $T_{mn}$ in (3.12). Varying $S_F$ $-$ with ${\mathcal L}_F$ in $S_F$ given by (4.25) where $V$ is now assumed to include the higher derivative terms $-$ with respect to $h^{mn}$ one finds that $T_{mn}$ in (3.12) becomes
\begin{equation}
T_{mn} = 2 (K_{mn}- \frac{1}{2} h_{mn} K\,)-2 \Big( \frac{\partial V}{\partial h^{mn}} - \frac{1}{2} h_{mn}V \Big) + u^{(1)}_{mn} \,\,,
\end{equation}
where $u^{(1)}_{mn} \equiv -2 {\mathcal D}_{mn}V $ and the functional derivative ${\mathcal D}_{mn}$ is defined by
\begin{equation}
{\mathcal D}_{mn} = h_{mk}h_{nl} \Bigg( \frac{1}{\sqrt{h_6}} \partial_a \Big({\sqrt{h_6}} \frac{\partial}{\partial (\partial_a h_{kl} )} \Big) - \frac{1}{\sqrt{h_6}} \partial_a \partial_b \Big({\sqrt{h_6}} \frac{\partial}{\partial (\partial_a \partial_b  h_{kl} )} \Big)\,\Bigg)\,\,.
\end{equation}
The nonzero values of $u^{(1)}_{mn}$ in (5.13) comes from the variations of the higher derivative terms like $\delta |{\mathcal R}_{abkl}|^2 / \delta h^{mn}$. Since $V$ now contains the Riemann and Ricci tensors in the string corrections, the variation $\delta \int d^6 y \sqrt{h_6}\, V / \delta h^{mn}$ generally contains the $u^{(1)}_{mn}$-type terms which includes the derivative ${\mathcal D}_{mn}$. Substituting (5.13) into (3.11) and contracting the indices $m$ and $n$ one obtains
\begin{equation}
{\mathcal R}_6 - {\mathcal L}_F - \frac{1}{2} ({\mathcal N}-1) V +\frac{u_{(1)}}{4} +\frac{3}{2} c \chi^{-1/2} =0 \,\,,
\end{equation}
where $u_{(1)} \equiv h^{mn} u^{(1)}_{mn}$.

Next, substituting ${\mathcal L}_F$ in (5.15) into (3.12) we get
\begin{equation}
T_{mn}= 2({\mathcal R}_{mn} - \frac{1}{2} h_{mn}{\mathcal R}_6 ) - \Big( \frac{\partial~~}{\partial h^{mn}} - \frac{1}{2} h_{mn} \Big)({\mathcal N}-1)V - \frac{1}{2} u^{(2)}_{mn} - \frac{3}{2} c \chi^{-1/2} h_{mn} \,\,,
\end{equation}
where $u^{(2)}_{mn}$ defined by
\begin{equation}
u^{(2)}_{mn} = - \Big( \frac{\partial ~~}{\partial h^{mn}} - \frac{1}{2} h_{mn} \Big) u_{(1)} + 2 {\mathcal D}_{mn} \Big(({\mathcal N}-1)V - \frac{u_{(1)}}{2} \Big) \,\,.
\end{equation}
Finally, substituting (5.16) into (3.11) and contracting $m$ and $n$ we obtain the modified version of (5.2):
\begin{equation}
c = - \frac{1}{3} \chi^{1/2} \Big(({\mathcal N}-1)({\mathcal N}-3)V + \frac{u_{(2)}}{2} \Big) \,\,,
\end{equation}
where ${u_{(2)}}$ is defined by ${u_{(2)}} \equiv h^{mn} u^{(2)}_{mn}$. The modified equation (5.18) reduces back to (5.2) if $V$ does not contain the higher derivative terms because in this case the second term ${u_{(2)}}/{2}$ does not exist anymore.

Before we end this section we examine the constituent equation (5.8) as a final check. First, the equation for ${\hat S}_{\rm bulk}$ in (4.30) was obtained by integrating (4.26) over the 6D internal space. But now (4.26) has been modified into (5.15). So we expect that the equation for ${\hat S}_{\rm bulk}$ is also modified by this modification of (5.15). Integrating (5.15) over the internal dimensions we obtain
\begin{equation}
{\hat S}_{\rm bulk} = - \frac{3c}{4 \kappa^2} - {\mathcal V}_{\rm scalar}^S \,\,,
\end{equation}
where ${\mathcal V}_{\rm scalar}^S$ is defined by
\begin{equation}
{\mathcal V}_{\rm scalar}^S = - \frac{1}{4 \kappa^2_{10} g_s^2} \int d^6 y \sqrt{h_6} \, \Big( ({\mathcal N}-1) V - \frac{u_{(1)}}{2} \Big)\,\,.
\end{equation}
${\hat S}_{\rm bulk}$ in (5.19) is essentially of the same form as ${\hat S}_{\rm bulk}$ in (4.30). The only change is that ${\mathcal V}_{\rm scalar}^S$ in (4.24) is generalized into (5.20). In the self-tuning mechanism of this paper the general definition of ${\mathcal V}_{\rm scalar}^S$ is given by (5.20). This definition coincides with (4.24) for $V \in V_3$, and also accords with the equation $\lambda = - \frac{\kappa^2}{2} {\mathcal V}_{\rm scalar}^S$ of the ordinary flex compactifications (see Eq.(5.8)).

Now we rewrite (5.19) as
\begin{equation}
{\hat S}_{\rm bulk} = - \frac{3}{4 \kappa^2} \Big(\frac{\beta}{\beta_Q}\Big)^2 \Big( 4 \lambda + \Delta {{\mathcal R}}_4 \Big) - {\mathcal V}_{\rm scalar}^S \,\,,
\end{equation}
and substitute (5.21) into (4.21). Then we obtain
\begin{equation}
\lambda = \frac{\kappa^2}{2} \Big(\frac{\beta_Q}{\beta}\Big)^2 \Big(- {\mathcal V}_{\rm scalar}^S + {\hat S}_{\rm brane} +{\hat S}_{\rm topological} \Big) - \frac{1}{8} \Delta {{\mathcal R}}_4 \,\,.
\end{equation}
The constituent equation (5.22) precisely coincides with the previous equation (5.8). As in the case of ${\hat S}_{\rm bulk}$, the constituent equation of $\lambda$ does not change its form either under the modification caused by ${\alpha^{\prime}}$-corrections. The above result implies that the equation of ${\hat S}_{\rm bulk}$, and the constituent equation of $\lambda$ are basically given, respectively, by the equations (5.19) and (5.22), where the scalar potential ${\mathcal V}_{\rm scalar}^S$ is most generally defined by (5.20). Since the constituent equation (5.22) in Sec. 5.3 precisely coincides with the equation (5.8) in Sec. 5.2, the self-tuning mechanism described in Sec. 5.2 also applies to the case with nonzero $\alpha^{\prime}$-corrections just as it is.

\vskip 1cm
\hspace{-0.65cm}{\bf \Large VI. Vanishing tree level actions}
\vskip 0.5cm
\setcounter{equation}{0}
\renewcommand{\theequation}{6.\arabic{equation}}

Now we show that the actions ${\hat S}_{\rm bulk}$, ${\hat S}_{\rm brane}$ and ${\hat S}_{\rm topological}$ all vanish at the tree level: i.e. we have
\begin{equation}
{\hat S}_{\rm bulk} ({\rm tree}) = {\hat S}_{\rm brane}({\rm tree}) ={\hat S}_{\rm topological} ({\rm tree}) =0 \,\,.
\end{equation}
Indeed, the vanishing of the tree level actions was already discussed in \cite{6} for the case $G_{(3)} =0$. Both ${\hat S}_{\rm bulk} ({\rm
tree})$ and ${\hat S}_{\rm brane} ({\rm tree})$ vanished by 6D Einstein equations and field equations for $\xi$ and $A$. In this section we will
show that this is also the case in the time-dependent models of this paper. ${\hat S}_{\rm brane}$ and ${\hat S}_{\rm bulk}$ (and even ${\hat
S}_{\rm topological}$) all vanish at the tree level even when $G_{(3)} \neq 0$.

\vskip 0.3cm
\hspace{-0.6cm}{\bf \large 6.1 6D field equations}
\vskip 0.15cm

We start with the 6D Einstein equation to show that ${\hat S}_{\rm bulk} ({\rm tree}) =0$ and ${\hat S}_{\rm brane}({\rm tree}) =0$.  A simple example of the type IIB configuration is given by the system which consists of the D3-beanes (and fractional D3-branes) located at the conifold singularity. In the following discussion we will take the conifold as our background manifold as in Sec. 3.5.  To understand the inside story of (6.1) more concretely we introduce a definite (but general) ansatz for the 6D metric\footnote{The metric (6.2) is equivalent to $ds^2_6 = \frac{dr^2}{f^2 (r)} + r^2 d \Sigma^2_{1,1}$, which is a typical ansatz for the $r$-dependent metric. The ansatz (6.2) has been introduced only for an easy understanding. In the following discussion we will only consider the case ${\mathcal R} (r) = r$.}
\begin{equation}
ds_6^2 \equiv h_{mn} ({y}) d {y}^m d {y}^n = d {r}^2 + R^2 (r) d \Sigma_{1,1}^2  \,\,
\end{equation}
as in \cite{6}, where
\begin{equation}
d \Sigma_{1,1}^2 = \frac{1}{9} \Big( d \psi + \sum_{i=1}^{2} \cos \theta_i d \phi_i \Big)^2 + \sum_{i=1}^{2} \frac{1}{6} \Big( d\theta_i^2 +
\sin^2 \theta_i d\phi_i^2 \Big)
\end{equation}
is an Einstein metric representing the base of the cone in the conifold metric. The 6D Einstein equations can be obtained from (3.7) as in (3.11).
For the metric (6.2) the 6D Einstein equations take the forms \cite{6}
\begin{equation}
{\mathcal L}_F + c \chi^{-1/2} - 20 \Big( \frac{{R^\prime}^2}{R^2 \,\,\,}  - \frac{1}{R^2} \Big) =0 \,\,,
\end{equation}
\begin{equation}
8\frac{R^{\prime\prime}}{R\,\,\,}+{\mathcal L}_F - c \chi^{-1/2} + 12 \Big( \frac{{R^\prime}^2}{R^2\,\,\,}  - \frac{1}{R^2} \Big) =0 \,\,,
\end{equation}
where the "prime" denotes the derivative with respect to $r$. (6.4) and (6.5) are the ${r}{r}$ and $\theta_i \theta_i$ components of the
Einstein equation and there is no any other independent equation besides these two.

Besides the above Einstein equations, we also need the equations for $\xi (r)$ and $\chi^{1/2} (r)$. The equations of these fields can be obtained
from the 6D total effective action $S_{\rm total}^{(6D)}$ defined by
\begin{equation}
S_{\rm total}^{(6D)} \equiv \big( S_{\rm IIB} + S_{\rm brane} \big)/ \Big({ \int d^4 {\hat x} \sqrt{-{\hat g}_4}\, \frac{\alpha^4}{\beta^4}
\Gamma_R^2 }\Big) \,\,,
\end{equation}
where $S_{\rm IIB}$ and $S_{\rm brane}$ are given by (3.7) and (4.4). Using (4.1) one can rewrite (6.6) as
\begin{equation}
S_{\rm total}^{(6D)}= \frac{1}{2 \kappa_{10}^2 g_s^2 } \int d^6 y \sqrt{h_6} \Big( {\mathcal R}_6 (h_{mn}) - {\mathcal L}_F + c \chi^{-1/2}
\Big) +{\rm topological ~term} \nonumber
\end{equation}
\begin{equation}
+ \int d^6 y \sqrt{h_6} \Big( - \frac{T_0}{g_s} \, \chi^{1/2} \, L (\gamma({t})) + \mu_0 \xi \Big) \delta^6 (y) \,\,,
\end{equation}
where ${\mathcal L}_F$ and the topological term are given by (3.8) and (3.10), respectively. $S_{\rm total}^{(6D)}$  in (6.7) precisely
coincides with the corresponding 6D effective action of the time-independent theory in \cite{3} if $L (\gamma({t})) =1$. In this section we are essentially looking for the classical (tree level) equations and therefore the tree level action in which $L(\gamma ({t}\,))$ is given by $L(\gamma ({t}\,)) =1$ is good enough to obtain the field equations we want. Now we can show that the field equation for $\xi$ obtained from (6.7) coincides with the time-independent equation (3.5). Also, varying (6.7) with respect to $A$ we obtain (see Sec. 7.2 of Ref. \cite{3})
\begin{equation}
\nabla^2 \Big( \frac{\chi^{1/2}}{g_s} \Big) = \frac{i}{12 Im \tau}\, \chi \, {\mathcal G}_{mnp} \big( \ast_6 {\bar {\mathcal G}}\big)^{mnp} +
\frac{1}{6 Im \tau} \, \chi \, {\mathcal G}_{mnp}^{+}{\bar {\mathcal G}}^{+ mnp} + \Big( \frac{\chi^{1/2}}{g_s} \Big)^{-1} \Big[ \partial \Big(
\frac{\chi^{1/2}}{g_s} \Big) \Big]^2 \nonumber
\end{equation}
\begin{equation}
+ \Big( \frac{\chi^{1/2}}{g_s} \Big)^{-1} (\partial \xi)^2 + \frac{c}{g_s} + 2 \kappa_{10}^2 T_0 \, \chi \, \delta^6 (y) \,\,.
\end{equation}

\vskip 0.3cm
\hspace{-0.6cm}{\bf \large 6.2 Vanishing tree level actions}
\vskip 0.15cm

Now we have four linearly independent equations (Eqs. (6.4), (6.5), (6.8) and (3.5)) defined on the 6D internal sector. In the following we will use these equations to show that the vanishing of tree level actions in (6.1) is really the case. Let us start with the 6D Einstein equations in (6.4) and (6.5). Since ${\mathcal R}(r) =r$ for the conifold metric, these two equations can be solved by
\begin{equation}
c=0\,\,,~~~~~~~~{\mathcal L}_F =0 \,\,,
\end{equation}
and from (6.9) one finds that ${\hat S}_{\rm bulk}$ in (4.3) vanishes at the tree level by the following reason. The background geometry of our internal space is given by the conifold metric $ds_{\rm conifold}^2 = dr^2 + r^2 d \Sigma_{1,1}^2$. Hence at the tree level ${\mathcal R}_6 (h_{mn})$ in (4.3) vanishes because the conifold metric is Ricci-flat. Also since ${\mathcal L}_F =0$ by (6.9), we find that ${\hat S}_{\rm bulk}$ in (4.3) vanishes at the tree level as mentioned above.

We turn to the equations ${\hat S}_{\rm brane} ({\rm tree}) =0$ and ${\hat S}_{\rm topological} ({\rm tree})=0$. We see that ${\hat S}_{\rm brane} ({\rm tree})$ in (4.7) can be rewritten (for $\mu_0 =T_0$) as
\begin{equation}
{\hat S}_{\rm brane} ({\rm tree}) = - \frac{T_0}{g_s} \, \int d^6 y \sqrt{h_6} \,\Phi_{-} (y) \,\delta^6 (y) \,\,,
\end{equation}
and similarly from (3.10)
\begin{equation}
{\hat S}_{\rm topological} =\frac{i}{4 \kappa_{10}^{2} {\rm Im} \tau}  \int \frac{\Phi_{-} (y)}{g_s} {\mathcal G}_{3} \wedge \bar {\mathcal G}_{(3)} \,\,,
\end{equation}
where $\Phi_{-}(y)$ is the D3-brane potential defined by (2.24). Now ${\hat S}_{\rm brane} ({\rm tree}) = 0$ and ${\hat S}_{\rm topological} ({\rm tree}) =0$ can be shown from the two remaining equations (3.5) and (6.8) as follows. Subtracting (3.5) from (6.8) (and also setting $\mu_0 =T_0$ again) one obtains
\begin{equation}
\nabla^2 \Phi_{-} = \frac{g_s}{6 Im \tau} \, \chi \, |{\mathcal G}_{(3)}^{+} |^2 + \chi^{-1/2}|\partial \Phi_{-}|^2 +c \,\,,
\end{equation}
where the last term $c$ will vanish by (6.9). Equation (6.12) shows that the IASD fluxes ${\mathcal G}_{(3)}^{+}$ become a source for the potential $\Phi_{-}$. But these IASD fluxes acquire nonzero values only at the quantum level and therefore ${\mathcal G}_{(3)}^{+}$ and $\Phi_{-}$ vanish in the ISD (i.e. tree level) background: ${\mathcal
G}_{-}^{(0)} =  \Phi_{-}^{(0)} =0$, where ${\mathcal G}_{-} \equiv - i {\mathcal G}_{(3)}^{+}$. (See Sec. 2.3  of Ref. \cite{8}.) So we find that (6.10) and (6.11) both vanish because they are proportional to $\Phi_{-}(y)$, and therefore we have ${\hat S}_{\rm brane} ({\rm tree}) ={\hat S}_{\rm topological} ({\rm tree}) = 0$ together with ${\hat S}_{\rm bulk} (\rm tree)=0$ as stated in (6.1).

We summarize the above results as follows. The tree level actions ${\hat S}_{\rm bulk} ({\rm tree})$, ${\hat S}_{\rm brane}$$({\rm tree})$ and
${\hat S}_{\rm topological} ({\rm tree})$ all vanish: ${\hat S}_{\rm bulk} ({\rm tree}) = {\hat S}_{\rm brane} ({\rm tree}) ={\hat S}_{\rm
topological} ({\rm tree})=0$ and therefore ${\hat S}_{\rm total}$ in (4.9) receives nonzero contributions only at the quantum level:
\begin{equation}
{\hat S}_{\rm total} = \delta_Q {\hat S}_{\rm bulk} + \delta_Q {\hat S}_{\rm brane} + \delta_Q {\hat S}_{\rm topological}  \,\,.
\end{equation}
In (6.13), $\delta_Q {\hat S}_{\rm brane}$ represents $\delta_Q {\hat S}_{\rm brane}^{(NS)} + \delta_Q {\hat S}_{\rm brane}^{(R)} - {\mathcal
E}_{\rm SB}$ as described in Sec. 5.2 and similarly $\delta_Q {\hat S}_{\rm bulk}$/$\delta_Q {\hat S}_{\rm topological}$ are the contributions
to ${\hat S}_{\rm bulk}$/${\hat S}_{\rm topological}$ coming from the quantum corrections. In the case of the bulk action $\delta_Q {\hat S}_{\rm bulk}$ can be written in terms of the scalar potential ${\mathcal V}_{\rm scalar}^S$ by (5.19). For instance, in the case of $c=0$, (5.19) reduces to
\begin{equation}
{\hat S}_{\rm bulk} = - {\mathcal V}_{\rm scalar}^S \,\,.
\end{equation}
So we have
\begin{equation}
\delta_Q {\hat S}_{\rm bulk} = - {\mathcal V}_{\rm scalar}^S \,\,,
\end{equation}
because ${\hat S}_{\rm bulk}(\rm tree) =0$ and therefore ${\hat S}_{\rm bulk} = \delta_Q {\hat S}_{\rm bulk}$. Originally, ${\hat S}_{\rm bulk}$ was defined by (4.3) and in our discussions we did not restricted it only to on-shell. Indeed, on the right hand side of (6.15) the main contributions to ${\mathcal V}_{\rm scalar}^S$ are off-shell contributions coming from perturbative and nonperturbative corrections. (Note that $V$ in ${\mathcal V}_{\rm scalar}^S$ is given by $\propto {\mathcal G}_{(3)}^{+} \cdot {\bar {\mathcal G}}_{(3)}^{+}$ in the absence of $\alpha^{\prime}$-corrections.) Collecting all these together, we find that ${\hat S}_{\rm total}$ in (6.13) can be written (in the case $c=0$) as
\begin{equation}
{\hat S}_{\rm total} = -{\mathcal V}_{\rm scalar}^S + \delta_Q {\hat S}_{\rm brane}^{(NS)} + \delta_Q {\hat S}_{\rm brane}^{(R)} + \delta_Q {\hat
S}_{\rm topological} - {\mathcal E}_{\rm SB} \equiv \delta_Q {\hat S}_{\rm total}\,\,,
\end{equation}
where we have used (5.12) for the definition of $\delta_Q {\hat S}_{\rm total}$.

\vskip 1cm
\hspace{-0.65cm}{\bf \Large VII. Fine-tunings of $\lambda$ and moduli stabilizations}
\vskip 0.5cm
\setcounter{equation}{0}
\renewcommand{\theequation}{7.\arabic{equation}}

\vskip 0.3cm
\hspace{-0.6cm}{\bf \large 7.1 Case of the low-energy supergravity ($c=0$)}
\vskip 0.17cm
\hspace{-0.45cm}{(1) Equations of motion for the scale factors}
\vskip 0.17cm

Equations of motion for the scale factors $\mathcal A$, $\beta$ and $\beta_Q$ can be obtained from the total action (4.19). (4.19) takes different forms according to whether we take the stringy effect (i.e. $\alpha^{\prime}$-corrections) into account or not. Note that the second term of (4.19) is given by ${\hat S}_{\rm total}$, and ${\hat S}_{\rm bulk}$ in ${\hat S}_{\rm total}$ contains a $c$-dependent term $- \frac{3c}{4 \kappa^2}$ (Eq. (5.19)). Since $c$ is proportional to ${\mathcal R}_4^{\rm (eff)}$ (Eq. (4.28)), this term can be added to the first term of (4.19) when $c \neq 0$, and then we obtain a different equation as compared with the case $c=0$. In the case of the low-energy supergravity this $c$ term vanishes because we have $c=0$ in this case (see Sec. 5.1). But once we admit $\alpha^{\prime}$-corrections, $c$ does not vanish anymore (we will see this in Sec. 7.2) and ${\hat S}_{\rm bulk}$ in (5.19) acquires the nonzero $c$ term which is to be added to the first term of (4.19) as mentioned above. In this section we first consider the case $c=0$, and then we turn to the case $c \neq 0$ in Sec. 7.2.

Using (4.14) and (4.17) one can rewrite (4.19) as
\begin{equation}
S_{\rm total} = \frac{1}{2\kappa^2} \int d^3 {\vec x} \int dt {\mathcal L}_{\rm total}  + {\rm surface~\,terms} \,\,,
\end{equation}
where the total Lagrangian ${\mathcal L}_{\rm total}$ is given by
\begin{equation}
{\mathcal L}_{\rm total} = {\mathcal A}^3 \Bigg(- 6 \Big(\frac{\dot{{\mathcal A}}}{{\mathcal A}} \Big)^2 + 2 \Big(\frac{\dot{\beta}_Q}{\beta_Q}
\Big)^2 + 6 \Big(\frac{\dot{\beta}}{\beta}\Big)^2 \Bigg) + 2 \kappa^2 {\mathcal A}^3 \Big(\frac{\beta_Q}{\beta}\Big)^2\, \delta_Q {\hat S}_{\rm total} \,\,,
\end{equation}
where we have used the relation ${\hat S}_{\rm total} = \delta_Q {\hat S}_{\rm total}$ in (6.16). The equations of motion following from (7.2) are
\begin{equation}
\frac{d}{dt} \Big(\frac{\dot{{\mathcal A}}}{{\mathcal A}} \Big) + \frac{3}{2}\Big(\frac{\dot{{\mathcal A}}}{{\mathcal A}} \Big)^2 +
\frac{1}{2}\Big( \frac{\dot{\beta}_Q}{\beta_Q} \Big)^2 + \frac{3}{2} \Big(\frac{\dot{\beta}}{\beta} \Big)^2 = -\frac{\kappa^2}{2}
\Big(\frac{\beta_Q}{\beta}\Big)^2\, \delta_Q {\hat S}_{\rm total} \,\,,
\end{equation}
\begin{equation}
\frac{d}{dt} \Big(\frac{\dot{\beta}_Q}{\beta_Q} \Big) + {3}\frac{\dot{{\mathcal A}}}{{\mathcal A}} \frac{\dot{\beta}_Q}{\beta_Q} = \kappa^2
\Big(\frac{\beta_Q}{\beta}\Big)^2  \, \delta_Q {\hat S}_{\rm total} \,\,,
\end{equation}
\begin{equation}
\frac{d}{dt}\Big(\frac{\dot{\beta}}{\beta} \Big) + 3 \frac{\dot{{\mathcal A}}}{{\mathcal A}} \frac{\dot{\beta}}{\beta} = - \frac{\kappa^2}{3}
\Big(\frac{\beta_Q}{\beta}\Big)^2 \, \delta_Q  {\hat S}_{\rm total} \,\,.
\end{equation}

Besides these equations, we also have a constraint equation ${\mathcal R}_4^{(\rm eff)} (g_{\mu\nu}, \beta_Q, \beta) =0$ which follows from $c=0$ (see (4.28) and (5.3)). Since ${\mathcal R}_4^{(\rm eff)} (g_{\mu\nu}, \beta_Q, \beta)$ is defined by (4.18), ${\mathcal R}_4^{(\rm eff)} (g_{\mu\nu}, \beta_Q, \beta) =0$ implies that
\begin{equation}
6 \frac{d}{dt} \Big(\frac{\dot{{\mathcal A}}}{{\mathcal A}} \Big) + 6 \frac{d}{dt} \Big(\frac{\dot{\beta}_Q}{\beta_Q} \Big) + 12
\frac{d}{dt}\Big(\frac{\dot{\beta}}{\beta} \Big) + 12 \Big(\frac{\dot{{\mathcal A}}}{{\mathcal A}} \Big)^2 + 2
\Big(\frac{\dot{\beta}_Q}{\beta_Q} \Big)^2 + 6 \Big(\frac{\dot{\beta}}{\beta}\Big)^2 +18 \frac{\dot{{\mathcal A}}}{{\mathcal A}}
\frac{\dot{\beta}_Q}{\beta_Q} + 36 \frac{\dot{{\mathcal A}}}{{\mathcal A}} \frac{\dot{\beta}}{\beta} =0 \,\,.
\end{equation}
Equation (7.6) is the last independent equation continued from the three equations in (7.3), (7.4) and (7.5). However, instead of (7.6), one can consider a different (i.e. a substitute) equation if he want. Namely a linear combination of (7.3), (7.4) and (7.5) gives
\begin{equation}
6 \frac{d}{dt} \Big(\frac{\dot{{\mathcal A}}}{{\mathcal A}} \Big) + 6 \frac{d}{dt} \Big(\frac{\dot{\beta}_Q}{\beta_Q} \Big) + 12
\frac{d}{dt}\Big(\frac{\dot{\beta}}{\beta} \Big) + 9 \Big(\frac{\dot{{\mathcal A}}}{{\mathcal A}} \Big)^2 + 3 \Big(\frac{\dot{\beta}_Q}{\beta_Q} \Big)^2 + 9 \Big(\frac{\dot{\beta}}{\beta}\Big)^2 +18 \frac{\dot{{\mathcal A}}}{{\mathcal A}} \frac{\dot{\beta}_Q}{\beta_Q} + 36
\frac{\dot{{\mathcal A}}}{{\mathcal A}} \frac{\dot{\beta}}{\beta} \nonumber
\end{equation}
\begin{equation}
= - \kappa^2 \Big(\frac{\beta_Q}{\beta}\Big)^2 \,
\delta_Q {\hat S}_{\rm total}\,\,,
\end{equation}
and subtracting (7.7) from (7.6) one obtains
\begin{equation}
3\Big(\frac{\dot{{\mathcal A}}}{{\mathcal A}} \Big)^2 - \Big(\frac{\dot{\beta}_Q}{\beta_Q} \Big)^2 -3 \Big(\frac{\dot{\beta}}{\beta}\Big)^2 =
\kappa^2 \Big(\frac{\beta_Q}{\beta}\Big)^2 \,\delta_Q  {\hat S}_{\rm total} \,\,,
\end{equation}
which can be used as a substitute for (7.6) in the following discussions.

Equation (7.8) is much simpler than (7.6) and convenient for the later analysis. By this replacement the set of 4D equations is now given by (7.3), (7.4), (7.5) and the substitute equation (7.8). We rewrite them below for reader's convenience.
\begin{equation}
2\frac{d}{dt} \Big(\frac{\dot{{\mathcal A}}}{{\mathcal A}} \Big) + 3\Big(\frac{\dot{{\mathcal A}}}{{\mathcal A}} \Big)^2 + \Big(
\frac{\dot{\beta}_Q}{\beta_Q} \Big)^2 + 3 \Big(\frac{\dot{\beta}}{\beta} \Big)^2 = - Q  \,\,,
\end{equation}
\begin{equation}
\frac{d}{dt} \Big(\frac{\dot{\beta}_Q}{\beta_Q} \Big) + {3}\frac{\dot{{\mathcal A}}}{{\mathcal A}} \frac{\dot{\beta}_Q}{\beta_Q} = Q \,\,,
\end{equation}
\begin{equation}
\frac{d}{dt}\Big(\frac{\dot{\beta}}{\beta} \Big) + 3 \frac{\dot{{\mathcal A}}}{{\mathcal A}} \frac{\dot{\beta}}{\beta} = -\frac{Q}{3}\,\,,
\end{equation}
\begin{equation}
3\Big(\frac{\dot{{\mathcal A}}}{{\mathcal A}} \Big)^2 - \Big(\frac{\dot{\beta}_Q}{\beta_Q} \Big)^2 -3 \Big(\frac{\dot{\beta}}{\beta}\Big)^2 =
Q \,\,,
\end{equation}
where $Q$ is a constant defined by
\begin{equation}
Q = \kappa^2 \Big(\frac{\beta_Q}{\beta}\Big)^2 \,\delta_Q {\hat S}_{\rm total} \,\,,
\end{equation}
and where $\beta$ and $\beta_Q$ represent their present values of our universe as before.

\vskip 0.17cm
\vskip 0.17cm
\hspace{-0.45cm}{(2) The fine-tuning $\lambda =0$ and fixed internal dimensions}
\vskip 0.17cm

Now we have a set of 4D equations (i.e. the equations from (7.9) to (7.12)) which must be satisfied by the present
values of $\frac{\dot{{\mathcal A}}}{{\mathcal A}}$, $\frac{\dot{\beta}_Q}{\beta_Q}$, $\frac{\dot{\beta}}{\beta}$ and their derivatives. To
solve these equations we introduce an usual ansatz for $\mathcal A (t)$ as
\begin{equation}
\mathcal A (t) = e^{Ht} \,\, ~~~~~~~~~\Big( H =\sqrt{\frac{\lambda}{3}}\,\Big) \,\,,
\end{equation}
where $H$ is the Hubble constant of the present universe and in the absence of $T_{\mu\nu}$ of the matter fields it is only given by $\lambda$ as in (7.14). Since $\lambda$ (and therefore $H^2$) of our universe is almost vanishing in the units of Planck density, $Q$ on the right hand sides of the 4D equations in (7.9) to (7.12) must also be almost vanishing because from (7.9) and (7.12) it is expected to be of an order $\sim \Big(\frac{\dot{{\mathcal A}}}{{\mathcal A}} \Big)^2 \sim \lambda$.

Now we introduce the ansatzs for
$\beta_Q$ and $\beta$ as
\begin{equation}
\beta_Q (t) = e^{q_{_0} t}\,\,, ~~~~~~ \beta(t) = e^{b_0 t}\,\,,
\end{equation}
which means that $q_{_0}$ and $b_0$ are the present values of $\frac{\dot{\beta}_Q}{\beta_Q}$ and $\frac{\dot{\beta}}{\beta}$. Since
$\frac{\dot{{\mathcal A}}}{{\mathcal A}}$ due to $\lambda$ is constant, the first term of (7.9) vanishes:\footnote{This is indeed the case when $T_{\mu\nu}$ (besides $\lambda g_{\mu\nu}$) vanishes. (See Sec. IV of \cite{9}, for instance.)}
\begin{equation}
\frac{d}{dt} \Big( \frac{\dot{A}}{A} \Big)  =0 \,\,,
\end{equation}
and the set of 4D equations from (7.9) to (7.12) reduces to
\begin{equation}
3H^2 + q_{_0}^2 + 3b_0^2 = -Q \,\,,
\end{equation}
\begin{equation}
d_Q + 3H q_{_0} = Q  \,\,,
\end{equation}
\begin{equation}
d_B + 3H b_0 = -\frac{Q}{3} \,\,,
\end{equation}
\begin{equation}
3H^2 - q_{_0}^2 - 3b_0^2 = Q \,\,,
\end{equation}
where
\begin{equation}
d_Q \equiv \frac{d}{dt} \Big( \frac{\dot{\beta}_Q}{\beta_Q} \Big) \,\,,~~~~~ d_B \equiv \frac{d}{dt} \Big( \frac{\dot{\beta}}{\beta} \Big)
\end{equation}
also represent their present ($t=0$) values of our universe, respectively.\footnote{To be precise, the ansatzs for $\beta_{Q} (t)$ and $\beta (t)$ may have to be written as $\beta_{Q} (t) = e^{{q_0 t} + \frac{1}{2} d_{Q} t^2 + \cdots}\,\,$ resp. $\beta(t) = e^{{b_0 t} + \frac{1}{2} d_{B} t^2 + \cdots}$ if $d_{Q}$ and $d_B$ in (7.21) take nonzero values. But in (7.15) the second (and higher) order terms $\frac{1}{2} d_{Q} t^2 + \cdots$ and $\frac{1}{2} d_{B} t^2 + \cdots$ in the exponents have been omitted because in the limit $H \cong 0$ they are quite negligible as compared with the first order terms ${q_0 t}$ and ${b_0 t}$ at around $t=0$. Note that $\frac{d_Q}{q_0}$ and $\frac{d_B}{b_0}$ are of an order $H$: $\frac{d_Q}{q_0} \sim H $ and $\frac{d_B}{b_0} \sim H $. See (7.53) and (7.54).}

Apart from this, one can show that $Q$ defined by (7.13) must be equal to $\lambda$. Eliminating $\Delta {\mathcal R}_4  (g_{\mu\nu}, \beta_Q, \beta)$ terms from (4.21) and (5.11) (and using ${\hat S}_{\rm total} = \delta_Q {\hat S}_{\rm total}$ of (6.16)) one can show that
\begin{equation}
 \lambda = Q \,\,.
\end{equation}
Now using $3H^2 = \lambda$ one finds that the solution satisfying (7.17) to (7.20) together with (7.22) is
\begin{equation}
\lambda =0 \,\,,~~~~~Q=0 \,\,,
\end{equation}
\begin{equation}
q_{_0}=0 \,\,, ~~~~~ b_0=0 \,\,,
\end{equation}
\begin{equation}
d_Q =0 \,\,, ~~~~~ d_B =0 \,\,.
\end{equation}
In the above solution $\lambda =0$ in (7.23) is consistent with $q_{_0}= b_0 =0$ in (7.24) by the following reason. First, $q_{_0}= b_0 =0$ means that $\frac{\dot{\beta}_Q}{\beta_Q}$ and $\frac{\dot{\beta}}{\beta}$ vanish and in this case $\Delta {\mathcal R}_4$ also vanishes because $\Delta {\mathcal R}_4$ is given by (4.22). Next, $\Delta {\mathcal R}_4 =0$ implies $\lambda =0$ by (5.6), so we see that $q_{_0}= b_0 =0$ implies $\lambda =0$ and therefore (7.23) is consistent with (7.24). Finally, one can check that $d_Q =d_B =0$ in (7.25) is required by (7.18) and (7.19).

The fine-tuning $\lambda =0$ (or $Q=0$) in (7.23) can be achieved by ${\mathcal E}_{\rm SB}$ in $\delta_Q {\hat S}_{\rm total}$ (see (5.12) or (6.16)). Since $\Delta {\mathcal R}_4$ vanishes in the present case, (5.11) reduces to
\begin{equation}
\lambda = \frac{\kappa^2}{2} \Big( \frac{\beta_Q}{\beta} \Big)^2 \delta_Q {\hat S}_{\rm total} \,\,,
\end{equation}
(see also (7.13)) and therefore the fine-tuning $\lambda =0$ (or $Q=0$) is equivalent to the requirement
\begin{equation}
\delta_Q {\hat S}_{\rm total} =0 \,\,,
\end{equation}
where $\delta_Q {\hat S}_{\rm total}$ is given by (5.12). But since $\delta_Q {\hat S}_{\rm total}$ includes ${\mathcal E}_{\rm SB}$ and this ${\mathcal E}_{\rm SB}$ has its own gauge arbitrariness, the quantum corrections $-{\mathcal V}_{\rm scalar}^S + \delta_Q {\hat S}_{\rm brane}^{(NS)} + \delta_Q {\hat S}_{\rm brane}^{(R)} + \delta_Q {\hat S}_{\rm topological}$ in $\delta_Q {\hat S}_{\rm total}$ can be compensated by ${\mathcal E}_{\rm SB}$ so that $\delta_Q {\hat S}_{\rm total}$, and therefore $\lambda$ vanishes as a result.

In any case, the solution given by (7.23), (7.24) and (7.25) can be rewritten as
\begin{equation}
\lambda =0, ~~~~~~~ \beta(t) = \beta_Q (t) = \gamma(t) = 1 \,\,,
\end{equation}
which means that the scale factor of the internal dimensions (and also the dilaton $e^{\Phi}$) does not evolve with time anymore. They remain fixed and at the same time the constraint equation $c =0$ reduces to $\lambda =0$. This result precisely coincides with the result of Ref. \cite{3}. However, though the results coincide, the stabilization mechanism of this paper is entirely distinguished from the mechanism used in \cite{3}. In \cite{3}, the scale factor of the internal dimensions (K$\ddot{\rm a}$hler modulus) is stabilized by a K$\ddot{\rm a}$hler modulus-dependent nonperturbative correction of KKLT and the no-scale structure of the scalar potential is broken as a result. But in this paper the K$\ddot{\rm a}$hler modulus is not stabilized by this KKLT scenario. The time-dependent scale factor of the internal dimensions is stabilized by a set of 4D equations defined on the external sector, which has nothing to do with the nonperturbative correction of the KKLT scenario. So the no-scale structure remains unbroken in the scenario of this paper and this provides a new type of stabilization mechanism distinguished from the conventional KKLT scenario.

\vskip 0.5cm
\hspace{-0.6cm}{\bf \large 7.2 Case with $\alpha^{\prime}$-corrections ($c \neq 0$)}
\vskip 0.17cm
\hspace{-0.45cm}{(1) Equations of motion for the scale factors}
\vskip 0.17cm

In Sec. 7.1 we obtained the fine-tuning $\lambda =0$ with fixed internal dimensions in the framework of the type IIB supergravity. In this case a set of 4D equations requires ${\beta}_Q (t) = \beta (t) =1$ together with $\lambda =0$. In the present section, we want to check the possibility of having nonzero $\lambda$ (and also having evolving internal dimensions as well) by considering the stringy effect of the string theory.

The full string theory requires the action (2.1) to admit $\alpha^{\prime}$-corrections that usually contain higher derivative terms (see for instance \cite{10,11}). These terms have many (contracted) indices and therefore do not belong to $V_n$ with $n=1$ or $3$. If we denote the collection of these terms by $\triangle V$, the scalar potential density now becomes
\begin{equation}
V=V_3 + \triangle V \,\,,
\end{equation}
where $V_3$ represents the ${\mathcal G}^+_{mnp} {\bar {\mathcal G}}^{+mnp}$ term which already exists in the supergravity Lagrangian ${\mathcal L}_F$. Since $V_3$ is projected out by the operator (${\mathcal N} -3$) in (5.2) (or (5.3)), it does not contribute to $c$ as we already know. But the correction terms in  $\triangle V$ do not satisfy this property. So in the presence of $\alpha^{\prime}$-corrections $c=0$ is not the case anymore. $c$ now takes nonzero values
\begin{equation}
c = - \frac{1}{3} \chi^{1/2} \Big(({\mathcal N}-1)({\mathcal N}-3) \triangle {V} + \frac{u_{(2)}}{2} \Big) \,\,,
\end{equation}
and as a result, (7.6) (which was obtained from (4.28) and $c=0$) should be corrected into
\begin{equation}
6\frac{d}{dt} \Big(\frac{\dot{{\mathcal A}}}{{\mathcal A}} \Big)+ 6\frac{d}{dt} \Big(\frac{\dot{\beta}_Q}{\beta_Q} \Big) +12\frac{d}{dt}\Big(\frac{\dot{\beta}}{\beta} \Big) +12 \Big(\frac{\dot{{\mathcal A}}}{{\mathcal A}} \Big)^2 + 2\Big(\frac{\dot{\beta}_Q}{\beta_Q} \Big)^2 + 6 \Big(\frac{\dot{\beta}}{\beta} \Big)^2 \nonumber
\end{equation}
\begin{equation}
+18 \frac{\dot{{\mathcal A}}}{{\mathcal A}} \frac{\dot{\beta}_Q}{\beta_Q} +36 \frac{\dot{{\mathcal A}}}{{\mathcal A}} \frac{\dot{\beta}}{\beta} =  \Big( \frac{\beta_Q}{\beta} \Big)^2 c \,\,.
\end{equation}

Besides this, there are also important changes in the equations (6.16) and (7.22). Since (5.19) does not imply (6.14) in the case $c \neq 0$, (6.15) should be changed into
\begin{equation}
\delta_Q {\hat S}_{\rm bulk} = -{\mathcal V}_{\rm scalar}^S - \frac{3c}{4 \kappa^2} \,\,,
\end{equation}
and therefore ${\hat S}_{\rm total} = \delta_Q {\hat S}_{\rm total}$ in (6.16) should also be changed into
\begin{equation}
{\hat S}_{\rm total} = \delta_Q {\hat S}_{\rm total} - \frac{3c}{4 \kappa^2} \,\,,
\end{equation}
in the case $c \neq 0$. So using (7.33), we find from (4.21) and (5.11) that (7.22) must be changed into
\begin{equation}
\lambda = Q - \frac{1}{4} \Big( \frac{\beta_Q}{\beta}\Big)^2 c \,\,.
\end{equation}

Now we go back to the 4D equations of motion in (7.3), (7.4) and (7.5). These equations were obtained from the Lagrangian (7.2). But in the case $c \neq 0$, ${\hat S}_{\rm total}$ is given by (7.33) instead of (6.16). So the Lagrangian (7.2) needs correction because it was obtained from the uncorrected equation (6.16). Using (7.33) and (4.28) we rewrite (4.19) as
\begin{equation}
S_{\rm total} = - \frac{1}{4 \kappa^2} \int d^3 {\vec x} \int dt {\mathcal L}_{\rm total} + {\rm surface ~\, term}\,\,,
\end{equation}
where ${\mathcal L}_{\rm total}$ is now given by
\begin{equation}
{\mathcal L}_{\rm total} = {\mathcal A}^3 \Bigg( -6 \Big(\frac{\dot{{\mathcal A}}}{{\mathcal A}} \Big)^2  +2\Big(\frac{\dot{\beta}_Q}{\beta_Q} \Big)^2 + 6 \Big(\frac{\dot{\beta}}{\beta} \Big)^2 \, \Bigg) - 4 \kappa^2 {\mathcal A}^3 \,\Big(\frac{\beta_Q}{\beta}\Big)^2\, \delta_Q {\hat S}_{\rm total} \,\,.
\end{equation}

The corrected Lagrangian (7.36) almost coincides with the original Lagrangian (7.2). It differs from (7.2) only in that $\delta_Q {\hat S}_{\rm total}$ in (7.2) is replaced by $-2 \delta_Q {\hat S}_{\rm total}$. So the equations following from (7.36) take the same forms as the original equations in (7.3), (7.4) and (7.5) only except that $\delta_Q {\hat S}_{\rm total}$ in the equations are replaced by $-2 \delta_Q {\hat S}_{\rm total}$. This prescription also applies to (7.7). The corrected version of (7.7) would be
\begin{equation}
6\frac{d}{dt} \Big(\frac{\dot{{\mathcal A}}}{{\mathcal A}} \Big)+ 6\frac{d}{dt} \Big(\frac{\dot{\beta}_Q}{\beta_Q} \Big) +12\frac{d}{dt}\Big(\frac{\dot{\beta}}{\beta} \Big) +9 \Big(\frac{\dot{{\mathcal A}}}{{\mathcal A}} \Big)^2 + 3\Big(\frac{\dot{\beta}_Q}{\beta_Q} \Big)^2 + 9 \Big(\frac{\dot{\beta}}{\beta} \Big)^2 ~~~~~~~~~~~~~~~~~~~~ \nonumber
\end{equation}
\begin{equation}
~~~~~~~~~ +18 \frac{\dot{{\mathcal A}}}{{\mathcal A}} \frac{\dot{\beta}_Q}{\beta_Q} +36 \frac{\dot{{\mathcal A}}}{{\mathcal A}} \frac{\dot{\beta}}{\beta} =  2 \kappa^2 \Big(\frac{\beta_Q}{\beta}\Big)^2 \delta_Q {\hat S}_{\rm total} \,\,,
\end{equation}
and subtracting (7.37) from (7.31) we obtain
\begin{equation}
3\Big(\frac{\dot{{\mathcal A}}}{{\mathcal A}} \Big)^2 - \Big(\frac{\dot{\beta}_Q}{\beta_Q} \Big)^2 -3 \Big(\frac{\dot{\beta}}{\beta} \Big)^2
= \Big( \frac{\beta_Q}{\beta} \Big)^2 c - 2 \kappa^2 \Big(\frac{\beta_Q}{\beta}\Big)^2  \delta_Q {\hat S}_{\rm total} \,\,,
\end{equation}
which is corrected version of the substitute equation (7.8).

Collecting all these together, we can now write the corrected versions of the 4D equations in Sec. 7.1 as
\begin{equation}
2\frac{d}{dt} \Big(\frac{\dot{{\mathcal A}}}{{\mathcal A}} \Big)+ 3 \Big(\frac{\dot{{\mathcal A}}}{{\mathcal A}} \Big)^2 + \Big(\frac{\dot{\beta}_Q}{\beta_Q} \Big)^2 + 3 \Big(\frac{\dot{\beta}}{\beta} \Big)^2  =2Q \,\,,
\end{equation}
\begin{equation}
\frac{d}{dt} \Big(\frac{\dot{\beta}_Q}{\beta_Q} \Big) + {3}\frac{\dot{{\mathcal A}}}{{\mathcal A}} \frac{\dot{\beta}_Q}{\beta_Q} = -2 Q \,\,,
\end{equation}
\begin{equation}
\frac{d}{dt}\Big(\frac{\dot{\beta}}{\beta} \Big) + 3 \frac{\dot{{\mathcal A}}}{{\mathcal A}} \frac{\dot{\beta}}{\beta} = \frac{2}{3}Q\,\,,
\end{equation}
\begin{equation}
3\Big(\frac{\dot{{\mathcal A}}}{{\mathcal A}} \Big)^2 - \Big(\frac{\dot{\beta}_Q}{\beta_Q} \Big)^2 -3 \Big(\frac{\dot{\beta}}{\beta}\Big)^2 =
-2Q  + \Big( \frac{\beta_Q}{\beta} \Big)^2 c \,\,.
\end{equation}

\vskip 0.17cm
\hspace{-0.45cm}{(2) Nonzero $\lambda$ and evolving internal dimensions}
\vskip 0.17cm

By the ansatzs (7.14) and (7.15) the equations from (7.39) to (7.42) reduce to
\begin{equation}
3H^2 + q_{_0}^2 + 3b_0^2 = 2Q \,\,,
\end{equation}
\begin{equation}
d_Q + 3H q_{_0} = -2Q  \,\,,
\end{equation}
\begin{equation}
d_B + 3H b_0 = \frac{2}{3}Q \,\,,
\end{equation}
\begin{equation}
3H^2 - q_{_0}^2 - 3b_0^2 = -2Q + \Big( \frac{\beta_Q}{\beta} \Big)^2 c \,\,,
\end{equation}
where the nonzero values of $c$ are given by (7.30), while $Q$ is still defined by (7.13). The above equations are the corresponding equations of (7.17), (7.18), (7.19) and (7.20) in Sec. 7.1, and similarly (7.34) of this section is the corresponding equation of (7.22) in Sec. 7.1.

Now we have five equations (Eqs. from (7.43) to (7.46) and Eq. (7.34)) which must be satisfied by the solution of this section. We solve these equations as follows. First, using $3H^2 = \lambda$ we recast the two equations in (7.43) and (7.46) into
\begin{equation}
\lambda = \frac{1}{2} \Big(\frac{\beta_Q}{\beta}\Big)^2 c \,\,,
\end{equation}
and
\begin{equation}
q_0^2 + 3b_0^2 = 2Q - \frac{1}{2} \Big(\frac{\beta_Q}{\beta}\Big)^2 c \,\,.
\end{equation}
Next, from (7.34) and (7.47) we obtain
\begin{equation}
Q= \frac{3}{4} \Big(\frac{\beta_Q}{\beta}\Big)^2 c \,\,,
\end{equation}
and therefore (7.47) and (7.48) can be rewritten as
\begin{equation}
\lambda = \frac{2}{3}Q \,\,,
\end{equation}
and
\begin{equation}
q_0^2 + 3b_0^2 = 2 \lambda \,\,.
\end{equation}
The constraint (7.49), or equivalently (7.50) can be achieved in any case by ${\mathcal E}_{\rm SB}$ contained in $Q$ (see (5.12) and (7.13)). Since ${\mathcal E}_{\rm SB}$ has gauge arbitrariness, $-{\mathcal V}_{\rm scalar}^S + \delta_Q {\hat S}_{\rm brane}^{(NS)} + \delta_Q {\hat S}_{\rm brane}^{(R)} + \delta_Q {\hat S}_{\rm topological}$ in $\delta_Q {\hat S}_{\rm total}$ can be gauged away by ${\mathcal E}_{\rm SB}$ so that $Q$ adjusts itself to satisfy (7.49), and therefore (7.50) as well.

The above result shows that the equations of this section admit solutions with nonzero $\lambda$. (7.47), which is the generalization of (7.23) to the case $c \neq 0$, shows that $\lambda$ takes nonzero values because $c$ in (7.30) does so. In addition, (7.51) requires that $\lambda$ must be positive because $q_0$ and $b_0$ are real. This positive $\lambda$, however, must be very small because it is of the same order as $c$ by (7.47). For instance, if $\triangle V$ in (7.30) is given by the effective Lagrangian in \cite{10}, then $c$, and therefore $\lambda$ must at least be of an order ${\alpha^{\prime}}^3$. Besides this, $c$ also contains the function $\chi^{1/2} (y)$. In the limit $G_{(3)} \rightarrow 0$ this function takes the form \cite{6}
\begin{equation}
\chi^{1/2} (r) = \Big( 1+ \frac{Q_0}{r^4} \Big)^{-1} \,\,, ~~~~~~~~~~ (Q_0 = {\rm constant})\,\,,
\end{equation}
and hence in the neighborhood of the brane at $r=0$ it reduces to $\chi^{1/2} (r) \sim r^4/Q_0$, which shows $c$ is highly suppressed again because $\chi^{1/2} (r)$ strongly vanishes at $r=0$ ($y=0$) in the approximation $G_{(3)} \cong 0$. These whole things imply that $\lambda$ in (7.47) should be very small anyway.

We finally determine $q_0$ and $b_0$ from (7.44), (7.45) and (7.51). Using (7.50) and $\lambda = 3H^2$ we rewrite them as
\begin{equation}
{\hat q}_0 = -3 - {\hat d}_Q \,\,,
\end{equation}
\begin{equation}
{\hat b}_0 = 1 - {\hat d}_B \,\,,
\end{equation}
\begin{equation}
{\hat q}_0^2 + 3 {\hat b}_0^2  = 6 \,\,,
\end{equation}
where ${\hat d}_Q$ and ${\hat d}_B$ are defined by
\begin{equation}
{\hat d}_Q \equiv \frac{d_Q}{3H^2} \,\,, ~~~~~{\hat d}_B \equiv \frac{d_B}{3H^2} \,\,,
\end{equation}
and similarly
\begin{equation}
{\hat q}_0 \equiv \frac{q_0}{H} \,\,, ~~~~~{\hat b}_0 \equiv \frac{b_0}{H} \,\,.
\end{equation}

The set of equations from (7.53) to (7.55) allows for infinite number of solutions. A unique and the most natural solution that comes to our mind from (7.53) and (7.54) would be obtained by considering an ansatz
\begin{equation}
{\hat q}_0 = -3{\hat b}_0 ~~~{\rm and}~~~  {\hat d}_Q = -3{\hat d}_B\,\,.
\end{equation}
The solution satisfying this ansatz is
\begin{equation}
{\hat q}_0 = -\frac{3}{\sqrt{2}} \,\,, ~~~~~{\hat b}_0 = \frac{1}{\sqrt{2}} \,\,,
\end{equation}
with
\begin{equation}
{\hat d}_Q = -3 \Big( 1-\frac{1}{\sqrt{2}} \Big) \,\,,~~~~~ {\hat d}_B = 1-\frac{1}{\sqrt{2}} \,\,.
\end{equation}
To see why this solution is unique, we go back to (4.10) and use (7.15) to rewrite it as
\begin{equation}
\gamma(t) = e^{(3 {\hat b}_0 + {\hat q}_0 ) Ht} \,\,.
\end{equation}
In (7.61), $\gamma(t)$ reduces to $\gamma(t) =1$ by (7.58), which means that the dilaton modulus is fixed
\begin{equation}
< e^{\Phi} > \,\,\,\rightarrow \,\,\, g_s  \,\,,
\end{equation}
despite that the scale factor $\beta(t)$ (the K$\ddot{\rm a}$hler modulus) still remains time-dependent in the above solution. Indeed we have
\begin{equation}
\beta(t) = e^{\frac{1}{\sqrt{2}} Ht}
\end{equation}
from (7.15) and (7.59), which suggests that the internal dimensions are now under accelerated expansion almost at the same rate as the external dimensions.

\vskip 1cm
\hspace{-0.65cm}{\bf \Large VIII. Summary and discussion}
\vskip 0.5cm
\setcounter{equation}{0}
\renewcommand{\theequation}{8.\arabic{equation}}

In this paper we presented a new scenario for the moduli stabilization and for a very small but nonzero positive $\lambda$ in the framework of the self-tuning mechanism proposed in \cite{3}. In our scenario the complex structure moduli are still stabilized by the three-form fluxes as in the usual flux compacifications. But the K$\ddot{\rm a}$hler modulus of the internal dimensions is not fixed by the KKLT-type mechanism. In our paper we assumed that the scale factor of the internal dimensions is basically allowed to evolve with time. But at the supergravity level the result of our scenario of this paper precisely coincides with the result of the time-independent theory presented in \cite{3}. We obtained $\lambda=0$, and fixed internal dimensions with $\beta(t)=1$.

Though the results coincide, the stabilization mechanism of this paper is very distinguished from the mechanism used in \cite{3}. In \cite{3}, the internal dimensions are stabilized by a K$\ddot{\rm a}$hler modulus-dependent nonperturbative corrections of KKLT and therefore the no-scale structure of the Lagrangian is broken in that case. But in this paper the internal dimensions are not stabilized by this KKLT scenario. The K$\ddot{\rm a}$hler modulus of the internal dimensions (and also the dilaton modulus) is stabilized by a set of 4D dynamical (plus consrtaint) equations defined on the external spacetime and the no-scale structure is unbroken in our scenario of this paper. In addition to all this, what is more important is that $\lambda$ is fine-tuned to zero, $\lambda =0$, at the level of supergravity approximation.

The above result changes once we admit $\alpha^{\prime}$-corrections of the string theory. Namely $\lambda=0$ changes into a new fine-tuning $\lambda = \frac{2}{3} Q$ (Eq. (7.50)), where $Q$ is related to the constant $c$ by the equation $Q = \frac{3}{4} \Big( \frac{\beta_Q}{\beta} \Big)^2 c$ (Eq. (7.49)), and where $c$ takes nonzero values arising from the $\alpha^{\prime}$-corrections (see (7.30)). So $\lambda$ in (7.50) acquires nonzero values from the $\alpha^{\prime}$-corrections and in the limit $\alpha^{\prime} \rightarrow 0$ it reduces back to $\lambda=0$ as it should be. But in any case, the fine-tunings of either $\lambda=0$ or $\lambda = \frac{2}{3} Q$ can be achieved by the supersymmetry breaking term ${\mathcal E}_{\rm SB}$ contained in $Q$ (see (7.13) and (5.12)). Namely, for a given value of $c$ (which is determined from the $\alpha^{\prime}$-corrections) ${\mathcal E}_{\rm SB}$ adjust itself (recall that ${\mathcal E}_{\rm SB}$ has gauge arbitrariness) so that $Q$ satisfies (7.49), and this nonzero $Q$ becomes a nonzero $\lambda$ by (7.50).

In the case of nonzero $\lambda$ (i.e. when we admit $\alpha^{\prime}$-corrections) the internal dimensions generically evolve with time as opposed to the case $\lambda=0$. Indeed the set of 4D equations requires that the scale factor of the internal dimensions(the K$\ddot{\rm a}$hler modulus)  must be of the form $\beta (t) = e^{{\hat b}_0 Ht}$, where ${\hat b}_0$ denotes dimensionless constants of order one. But since $H \cong 0$, $\beta (t)$ is virtually constant, $\beta (t) \cong 1$, and this means that the K$\ddot{\rm a}$hler modulus is almost fixed even in the case with $\alpha^{\prime}$-corrections. Among these solutions, of particular interest is the one that given by $\beta(t) = e^{\frac{1}{\sqrt{2}} Ht}$ (Eq. (7.63)). This solution corresponds to $\gamma(t)=1$, which means that the string coupling $e^{\Phi}$ of this solution remains time-independent (i.e. the dilaton modulus is fixed) though the internal dimensions evolve with time. Except for this particular case, $\beta(t)$ and $\gamma(t)$ are generically nontrivial (exponential) functions of time in the case of nonzero $\lambda$. But even in this case the moduli $\beta(t)$ and $\gamma(t)$ are stabilized in the sense that they are virtually constants, i.e. $\beta(t)$, $\gamma(t) \cong 1$, in the limit $H \cong 0$.

We finally discuss the observational effects of the time-evolving $\beta(t)$ and $\gamma(t)$. Since $\beta (t)$ is given by $\beta (t)= e^{{\hat b}_0 Ht}$ with ${\hat b}_0$ being of order one, the internal dimensions must expand almost at the same rate as the 4D external spacetime and such an expansion of the internal dimensions may lead to time-varying constants of nature. For instance, it is well known \cite{12} that the coupling constants $g_c$ including electric charges are inversely proportional to the radius $R$ of the compact internal dimensions:
\begin{equation}
g_c \sim \frac{\kappa}{R} \,\,,
\end{equation}
where $\kappa \equiv \sqrt{16 \pi G}$. Also in string theory the 4D gravitational constant $G$ behaves like \cite{13}
\begin{equation}
G \sim \alpha^{\prime} e^{2 \Phi} \,\,.
\end{equation}
So the coupling constants $g_c (t)$ take the form
\begin{equation}
g_c (t) \sim g_s \, {\alpha^{\prime}}^{1/2}\, \frac{e^{\Phi_S - \frac{B}{2}}}{R_0}\,\frac{\gamma(t)}{\beta(t)} \,\,,
\end{equation}
and for the solution with $\beta(t) = e^{\frac{1}{\sqrt 2} H t}$ and $\gamma(t)=1$ they become
\begin{equation}
g_c (t) \sim  g_c (0) e^{-\frac{1}{\sqrt 2} H t} \,\,,
\end{equation}
where $g_c (0) \equiv g_s \,{\alpha^{\prime}}^{1/2} \, \frac{e^{\Phi_S - \frac{B}{2}}} {R_0}$ and $R_0$ is a constant representing characteristic scale (radius) of the internal dimensions.

The above result suggests that some of the well-known constants of nature might not be real constants. For instance, (8.2) shows that the gravitational constant $G$ is proportional to $\gamma^2 (t)$. So if we choose the solutions with ${\dot \gamma} (t) \neq 0$, then $G$ becomes a time-varying quantity.\footnote{But if we choose $\gamma(t) =1$, then $G$ is still real constant.} Similarly, (8.4) shows that the coupling constants including electric charges are also time-varying quantities. They are decreasing in magnitudes at the same rate as the expansions of the internal and external dimensions. The decreasing or expanding rate of these quantities is very small. It is about $\sim e^{Ht}$, where the Hubble constant $H$ of our present universe is roughly given by \cite{14}
\begin{equation}
H^{-1} \sim 10^{10}\, {\rm yr} \,\,.
\end{equation}
So, for instance, the electric charges of our present universe decrease in magnitudes at the rate in which they become half the original magnitudes during about $10^{10}$ years, while the internal dimensions become doubled in size during that time. If these results can be checked by experiments, then it would be great. But if not, then we go back to Ref. \cite{3} to see what happens to the self-tuning equation in \cite{3} when we admit $\alpha^{\prime}$-corrections.

\vskip 1cm
\begin{center}
{\large \bf Acknowledgement}
\end{center}

This research was supported by Basic Science Research Program through the National Research Foundation of Korea (NRF) funded by the Ministry of Education (Grant No. 2018R1D1A1B07050146).

\vskip 1cm
\setcounter{equation}{0}
\renewcommand{\theequation}{A.\arabic{equation}}

\hspace{-0.6cm}{\bf \Large F-term scalar potential in Einstein frame}
\vskip 0.15cm

In the Einstein frame the type IIB 10D action is given by
\begin{equation}
S_{\rm IIB}^{\rm E} = \frac{1}{2 \kappa_{10}^2} \int d^{10}x \sqrt{-G_{10}^{\rm E}} \Big( {\mathcal R}_{10}^{\rm E} - \frac{1}{2} \partial_M \Phi \partial^M \Phi  - \frac{1}{2
\cdot 3!} e^{\Phi} G_{(3)} \cdot {\bar{G}_{(3)}} - \frac{1}{4 \cdot 5!} {\tilde F}_{(5)}^2  \Big) \nonumber
\end{equation}
\begin{equation}
~~~~~~~~~~+ \frac{1}{8 i \kappa_{10}^2} \int e^{\Phi} A_{(4)} \wedge G_{(3)} \wedge {\bar G}_{(3)}  \,\,,
\end{equation}
where we have set $A_{(0)} = 0$, and therefore $F_{(1)} = 0$ as in (2.1). The type IIB string frame metric (2.2) is related to the Einstein metric $ds_{\rm E}^2$ by the equation $ds^2 = e^{\frac{1}{2} \Phi}ds_{\rm E}^2$, so $ds_{\rm E}^2$ can be written as
\begin{equation}
ds_{\rm E}^2 =\alpha_{\rm E}^2 ({\hat t}) e^{A_{\rm E}(y)} {\hat g}_{\mu\nu} ({\hat x}) d{\hat x}^{\mu} d{\hat x}^{\nu} + \beta_{\rm E}^2 ({\hat t}) e^{B_{\rm E}(y)} {h}_{mn} (y)
d {y}^m d {y}^n \,\,,
\end{equation}
where $\alpha_{\rm E}^2 ({\hat t})$, $\beta_{\rm E}^2 ({\hat t})$ and $A_{\rm E}(y)$, $B_{\rm E}(y)$ are given, respectively, by
\begin{equation}
\alpha_{\rm E}^2 ({\hat t}) = \frac{1}{\sqrt{g_s}} \frac{\alpha^2 ({\hat t})}{\sqrt{\gamma ({\hat t})}} \,\,,~~~~~ \beta_{\rm E}^2 ({\hat t}) = \frac{1}{\sqrt{g_s}} \frac{\beta^2 ({\hat t})}{\sqrt{\gamma ({\hat t})}} \,\,,
\end{equation}
and
\begin{equation}
A_{\rm E}(y) = A(y) - \frac{\Phi_S (y)}{2} \,\,,~~~~~ B_{\rm E}(y) = B(y) - \frac{\Phi_S (y)}{2} \,\,,
\end{equation}
and since $B(y) = \Phi_S (y) - A(y)$, $A_{\rm E}(y)$ and $B_{\rm E}(y)$ satisfy $B_{\rm E}(y) = -A_{\rm E}(y)$.

Now $S_{\rm IIB}^{\rm E}$ in (A.1) can be reduced by (A.2). Among the terms in the reduced action, the four-dimensional curvature and ${\mathcal G}_{mnp}^+ {\bar{\mathcal G}}^{+mnp}$ terms are particularly important in the present discussion of this section. They are given by
\begin{equation}
 S_{\rm IIB}^{\rm E} = \frac{1}{2 \kappa_{10}^2} \bigg( \int d^4 {\hat x} \sqrt{-{\hat g}_4} \, \alpha_{\rm E}^2 \, \beta_{\rm E}^6 \, \hat{{\mathcal R}}_4^{\rm{E(eff)}} ({\hat g}_{\mu\nu}, \alpha_{\rm E}, \beta_{\rm E} )\bigg) \bigg( \int d^6  y \sqrt{h_6}\, e^{-2A_{\rm E}}  \bigg)   \nonumber
\end{equation}
\begin{equation}
 - \frac{1}{12 \kappa_{10}^2 } \bigg( \int d^4 {\hat x} \sqrt{-{\hat g}_4} \, \alpha^4_{\rm E} \eta^2 \bigg) \bigg(\int d^6 y \sqrt{h_6} \, e^{2A_{\rm E}} \, \frac{{\mathcal G}_{mnp}^+ {\bar{\mathcal G}}^{+mnp}}{{\rm Im} \tau} \bigg) + \cdots \,\,,
\end{equation}
where $\hat{{\mathcal R}}_4^{\rm E(eff)} ({\hat g}_{\mu\nu}, \alpha_{\rm E}, \beta_{\rm E})$ is the four-dimensional effective curvature scalar defined by (2.8) with ($\alpha ({\hat t})$, $\beta ({\hat t})$, $\gamma ({\hat t})$) replaced by ($\alpha_{\rm E} ({\hat t})$, $\beta_{\rm E} ({\hat t})$, $1$). Also the four-dimensional metric ${\hat g}_{\mu\nu}  ({\hat x})$ is still given by (2.3),
\setcounter{equation}{2}
\renewcommand{\theequation}{2.\arabic{equation}}
\begin{equation}
 {\hat g}_{\mu\nu} ({\hat x}) d{\hat x}^{\mu} d{\hat x}^{\nu} = - d {\hat t}^2 + a^2 ({\hat t})  d \vec{x}^2_3 \,\,.
\end{equation}
\setcounter{equation}{5}
\renewcommand{\theequation}{A.\arabic{equation}}
Now we make a coordinate transformation ${\hat t} \rightarrow t$ defined by
\begin{equation}
dt = \alpha_{\rm E} \beta_{\rm E}^3 d{\hat t}   \,\,,
\end{equation}
and introduce a new scale factor ${\mathcal A}(t)$ given by
\begin{equation}
{\mathcal A} = a \alpha_{\rm E}  \beta_{\rm E}^3  \,\,.
\end{equation}
In this new coordinate system the Einstein frame metric (A.2) changes into
\begin{equation}
ds_{\rm E}^2 =\frac{1}{\beta_{\rm E}^6 (t)} e^{A_{\rm E}(y)} {g}_{\mu\nu}(x) dx^{\mu} d x^{\nu} + \beta_{\rm E}^2 (t) e^{-A_{\rm E}(y)} {h}_{mn} (y)
d {y}^m d {y}^n \,\,,
\end{equation}
where the four-dimensional metric $g_{\mu\nu}(x)$ is now given by
\begin{equation}
 g_{\mu\nu}(x) dx^{\mu} dx^{\nu} = - dt^2 + {\mathcal A}^2 (t) d \vec{x}^2_3 \,\,.
\end{equation}
Also the two terms in (A.5) appear (refer to Sec. 4.2 for the first term) in the form
\begin{equation}
 S_{\rm IIB}^{\rm E} = \frac{1}{2 \kappa^2} \int d^3\vec{x} \int dt \sqrt{-g_4} \, {{\mathcal R}}_4^{\rm E(eff)}({g}_{\mu\nu}, \alpha_{\rm E}, \beta_{\rm E}) +  \int d^3\vec{x} \int dt \sqrt{-g_4} \, {\hat S}_{\rm G} + \cdots \,\,,
\end{equation}
where $2 \kappa^2 = {2 \kappa_{10}^2} / (\int d^6  y \sqrt{h_6} e^{-2A_{\rm E}})$ and ${\hat S}_{\rm G}$ is defined by
\begin{equation}
{\hat S}_{\rm G} = - \frac{1}{2 \kappa_{10}^2} \frac{\eta^2}{\beta^{12}_{\rm E}} \int_{\mathcal M} d^6 {y} \sqrt{h_6} \,e^{2A_{\rm E}} \, \frac{{\mathcal G}_{mnp}^+ {\bar{\mathcal G}}^{+mnp}}{{\rm Im} \tau} \equiv - {\mathcal V}_{\rm scalar} \,\,.
\end{equation}
${\mathcal V}_{\rm scalar}$ in (A.11) will be identified as the F-term scalar potential of the $N=1$, $D=4$ supergravity. We have
\begin{equation}
{\mathcal V}_{\rm scalar} =  - \frac{1}{2 \kappa_{10}^2}  \frac{1}{{\rm Im} \tau}  \frac{\eta^2}{\beta^{12}_{\rm E}} \int_{\mathcal M} \,e^{2A_{\rm E}} \, {\mathcal G}_{(3)}^{+} \wedge \ast_{6} {\bar {\mathcal G}}_{(3)}^{+} \,\,.
\end{equation}


\newpage
\vskip 1cm

\begin{thebibliography}{999}

\bibitem{1} S. Perlmutter et al., {\it Measurement of $\Omega$ and $\Lambda$ from 42 high-redshift supernovae}, Astrophys. J. {\bf 517} (1999) 565 ; A. G. Riess et al., {\it Observational Evidence from Supernovae for an Accelerating Universe and a Cosmological Constant}, Astron. J. {\bf 116} (1998) 1009 ; {\it BVRI Light Curves for 22 Type Ia Supernovae}, Astron. J. 117, 707 (1999) [arXiv:astro-ph/9810291].

\bibitem{2} S. Weinberg, {\it The cosmological constant problem}, Rev. Mod. Phys. 61 (1989) 1.

\bibitem{3} E. K. Park and P. S. Kwon, {\it Towards the core of the cosmological constant problem}, PTEP {\bf 2016} (2016) 013B05 [arXiv:1404.4437].

\bibitem{4} S. Kachru, R. Kallosh, A. Linde and S. P. Trivedi, {\it de Sitter Vacua in String Theory}, Phys. Rev. {\bf D68} (2003) 046005
    [arXiv:hep-th/0301240].

\bibitem{5} S. B. Giddings, S. Kachru and J. Polchinski, {\it Hierarchies from Fluxes in String Compactifications}, Phys. Rev. {\bf D66} (2002) 106006 [arXiv:hep-th/0105097].

\bibitem{6} E. K. Park and P. S. Kwon, {\it Remark on Calabi-Yau vacua of the string theory and the cosmological constant problem}, Phys. Rev.
    {\bf D88} (2013) 046007 [arXiv:1301.1783].

\bibitem{7} S. Kachru, M. Schulz and E. Silverstein, {\it Self-tuning flat domain walls in 5d gravity and string theory}, Phys. Rev. D62 (2000) 045021 [arXiv:hep-th/0001206].

\bibitem{7-1} A. Strominger, {\it Superstrings with torsion}, Nucl. Phys. {\bf B274} (1986) 253; C.M. Hull, {\it Compactifications of the Heterotic Superstring}, Phys. Lett. {\bf B178} (1986) 357.

\bibitem{8} D. Baumann, A. Dymarsky, S. Kachru, I. R. Klebanov and L. McAllister, {\it $D3$-brane Potentials from Fluxes in AdS/CFT}, [arXiv:1001.5028].

\bibitem{8-1} See, for instance, K. Becker, M. Becker and J. H. Schwarz, {\it String Theory and M-Theory} ; Cambridge University Press (2006).

\bibitem{9} E. J. Copeland, M. Sami and S. Tsujikawa, {\it Dynamics of dark energy}, Int. J. Mod. Phys. {\bf D15} (2006) 1753 [arXiv:hep-th/0603057].

\bibitem{10} M.D. Freeman, C.N. Pope, M.F. Sohnius and K.S. Stelle, {\it Higher-order $\sigma$-model counterms and the effective action for superstringsOriginal Research Article}, Phys. Lett. {\bf B178} (1986) 199 ; M. T. Grisaru and D. Zanon, {\it Sigma-model superstring corrections to the Einstein-Hilbert actionOriginal Research Article}, Phys. Lett. {\bf B177} (1986) 347.

\bibitem{11} D. J. Gross and Edward Witten, {\it Superstring modifications of Einstein's equations}, Nucl. Phys. {\bf B277} (1986) 1 ; I. Antoniadis, S. Ferrara, R. Minasian and K.S. Narain, {\it $R^4$ Couplings in M and Type II Theories on Calabi-Yau spaces}, Nucl. Phys. {\bf B507} (1997) 571 [arXiv:hep-th/9707013] ; S. Frolov, I. R. Klebanov and A. A. Tseytlin, {\it String Corrections to the Holographic RG Flow of Supersymmetric $SU(N) \times SU(N+M)$ Gauge Theory}, Nucl. Phys. {\bf B620} (2002) 84 [arXiv:hep-th/0108106] ; K. Becker, M. Becker, M. Haack, and J. Louis, {\it Supersymmetry breaking and $\alpha^{\prime}$-corrections to flux induced potentials}, J. High Energy Phys. {\bf 06} (2002) 060 [arXiv:hep-th/0204254].

\bibitem{12} See, for instance, A. Chodos and S. Detweiler, {\it Where has the fifth dimension gone?}, Phys. Rev. {\bf D21} (1980) 2167 ; S. Weinberg, {\it Charges from extra dimensions}, Phys. Lett. {\bf B125} (1983) 265.

\bibitem{13} B. Zwiebach, {\it A first course in string theory} ; Cambridge University Press (2004).

\bibitem{14} See, for instance, E. W. Kolb and M. S. Turner, {\it The early universe} ; Addison Wesley Publishing Company (1990).

\end{thebibliography}
\end{document}